\documentclass[12pt]{article}
\usepackage{amssymb}
\input  epsf  
\def\rlx{\relax\leavevmode}
\def\inbar{\vrule height1.5ex width.4pt depth0pt}
\def\IZ{\rlx\hbox{\small \sf Z\kern-.4em Z}}
\def\IR{\rlx\hbox{\rm I\kern-.18em R}}
\def\ID{\rlx\hbox{\rm I\kern-.18em D}}
\def\IC{\rlx\hbox{\,$\inbar\kern-.3em{\rm C}$}}
\def\IN{\rlx\hbox{\rm I\kern-.18em N}}
\def\IP{\rlx\hbox{\rm I\kern-.18em P}}
\def\one{\hbox{{1}\kern-.25em\hbox{l}}}

\def\beq{\begin{equation}}
\def\eeq{\end{equation}}
\def\bea{\begin{eqnarray}}
\def\eea{\end{eqnarray}}
\def\ber{\begin{array}}
\def\eer{\end{array}}

\begin{document}

\begin{titlepage}

November 2002 \hfill {UTAS-PHYS-02-03}\\

\vskip 1.6in
\begin{center}
{\Large {\bf Polynomial super-$gl(n)$ algebras }}
\end{center}

\normalsize
\vskip .4in

\begin{center}
P D Jarvis$^{\dagger}$ and G Rudolph$^{\dagger *}$,
 
\par \vskip .1in \noindent
{\it School of Mathematics and Physics, University of Tasmania}\\
{\it GPO Box 252-21, Hobart Tas 7001, Australia }\\

\end{center}
\par \vskip .3in \noindent

\vspace{1cm} We introduce a class of finite dimensional nonlinear 
superalgebras $L = L_{\overline{0}} + L_{\overline{1}}$ providing 
gradings of $L_{\overline{0}} = gl(n) \simeq sl(n) + gl(1)$.  Odd 
generators close by anticommutation on polynomials (of degree $>1$) in 
the $gl(n)$ generators.  Specifically, we investigate `type I' 
super-$gl(n)$ algebras, having odd generators transforming in a single 
irreducible representation of $gl(n)$ together with its 
contragredient.  Admissible structure constants are discussed in terms 
of available $gl(n)$ couplings, and various special cases and 
candidate superalgebras are identified and exemplified via concrete 
oscillator constructions.  For the case of the $n$-dimensional 
defining representation, with odd generators $Q_{a}, \; 
\overline{Q}\mbox{}^{b}$, and even generators ${E^{a}}_{b}$, $a,b = 
1,\ldots,n$, a three parameter family of quadratic super-$gl(n)$ 
algebras (deformations of $sl(n/1)$) is defined.  In general, 
additional covariant Serre-type conditions are imposed, in order that 
the Jacobi identities be fulfilled.  For these quadratic super-$gl(n)$ 
algebras, the construction of Kac modules, and conditions for 
atypicality, are briefly considered.  Applications in quantum field 
theory, including Hamiltonian lattice QCD and space-time 
supersymmetry, are discussed.

\vfill
\rule{8cm}{.2mm} \\
\footnotesize{ $^{\dagger}$Alexander von Humboldt Fellow \\ $^{*}$Institute for Theoretical Physics, University of Leipzig,
Augustusplatz 10-11, D-0419 Leipzig, Germany, \hspace{.2cm}
\texttt{rudolph@rz.uni-leipzig.de}}
\end{titlepage}
\section*{\S 1 Introduction}

The interplay between the application of symmetry principles to models
of physical systems, and study of the classification, properties and
representation theory of underlying algebraic structures, has long
been a major theme in mathematical physics.  A broad spectrum of
generalised symmetry algebras is under active study, including
infinite dimensional algebras and superalgebras, deformations of
universal enveloping algebras, and various ternary and other
non-associative algebras.  A particular class is the
so-called $W$-algebras and superalgebras, arising from Hamiltonian
reduction of systems with first class constraints defined on
Lie-Poisson manifolds.  Although the general study of non-linear Lie
(super-)algebras belongs to abstract deformation theory, in specific
contexts enough structure exists to allow progress on classification
and representation theory.  For example, the
$W$-(super)algebras, although not Lie (super)algebras, are
rigidly constrained by their origins in Hamiltonian reduction.

In this spirit we study in this paper a class of finite dimensional
nonlinear superalgebras, with attention to their covariant closure
relations, and which are defined algebraically, without reference to
additional structure.  Namely, we consider superalgebras (${\mathbb
Z}_{2}$-graded algebras) $L =
L_{\overline{0}}+L_{\overline{1}}$ with even subalgebra
$L_{\overline{0}}$, and odd subalgebra $L_{\overline{1}}$, with
defining relations of the form:
\begin{equation}
{[}L_{\overline{0}}, L_{\overline{0}}{]} \subseteq L_{\overline{0}}, 
\quad 
{[}L_{\overline{0}}, L_{\overline{1}}{]} \subseteq L_{\overline{1}}, \quad
{\{}L_{\overline{1}}, L_{\overline{1}}{\}} \subseteq U(L_{\overline{0}}).
\label{eq:GeneralDefL0L1}
\end{equation}
Such `nonlinear super-$L_{\overline{0}}$ algebras' possess odd
generators $L_{\overline{1}}$ whose anticommutation relations
generalise the defining relations of Lie superalgebras, in that they
close only in the universal enveloping algebra $U(L_{\overline{0}})$, 
that is, on \textit{polynomials} (of quadratic or higher
degree) in the even generators $L_{\overline{0}}$.  In this work we
take the latter to be the classical Lie algebra $L_{\overline{0}} =
gl(n) \simeq sl(n) + gl(1)$.

Study of the classification of such superalgebras devolves to 
examination of possible $L_{\overline{0}}$-modules $L_{\overline{1}}$, 
and admissible structure constants (\ref{eq:GeneralDefL0L1}) which are 
consistent with the Jacobi identities.  This is taken up in \S 2 
below, where we discuss the structure of candidate polynomial 
super-$gl(n)$ algebras of `type I': that is, where $L_{\overline{1}}$ 
consists of the direct sum of an (arbitrary) irreducible 
$L_{\overline{0}}$-module $\{ \lambda \}$ together with its 
contragredient representation $\{\overline{\lambda} \}$, denoted here 
$gl_{k}(n/ \{ \lambda \} + \{ \overline{\lambda} \} )$ (where $k$ is 
the maximal degree within $U(L_{\overline{0}})$ of the polynomials ${[} 
L_{\overline{1}}, L_{\overline{1}} {]}$).  In general, many types of 
structure constant are in principle allowed (identified as tensor 
couplings).  These must be enumerated in specific cases, and the 
Jacobi identities imposed in order to identify viable solutions.  
Examples include odd generators in totally antisymmetric and totally 
symmetric tensor representations, together with their contragredients.  
In \S 3 concrete low rank examples of this type are provided via 
explicit oscillator constructions, together with generalisations 
including parafermionic realisations.

The examples of \S 3 fulfil the desired anticommutation relations 
because of structure specific to the oscillator realisations.  In \S 4 
a more complete treatment is given, for one special case, by 
examination of the quadratic super-$gl(n)$ algebras, with even 
generators ${E^{a}}_{b}$, $ 1 \le a,b \le n$ in the Gel'fand basis, 
and odd generators $Q_{a}$, $\overline{Q}\mbox{}^{b}$ in the 
\textit{defining} $n$-dimensional representation of $gl(n)$, and its 
contragredient.  A three-parameter family of quadratic algebras 
$gl_{2}(n/ \{ 1 \} + \{ \overline{1} \} )^{\texttt{a}, \alpha,\beta}$ 
is identified, which closely parallels the well-known \textit{linear} 
super-$gl(n)$ algebra, namely the simple Lie superalgebra $sl(n/1)$ 
$\equiv gl_{1}(n/ \{ 1 \} + \{ \overline{1} \} )$.  In general, the 
Jacobi identities are satisfied provided additional covariant 
Serre-type relations of the form ${E^{a}}_{b}\overline{Q}\mbox{}^{b} = 
q \overline{Q}\mbox{}^{a}$, $ Q_{a}{E^{a}}_{b} = Q_{b} q $ hold in the 
enveloping algebra, for some $gl(n)$ invariant $q = \alpha \langle E 
\rangle + \beta \one$, where $\langle E \rangle = {E^{c}}_{c}$.  In 
\S 4 an outline of the construction of Kac modules is also given, 
together with the derivation of a necessary condition for typicality.

In the concluding remarks (\S 5 below), additional motivation for the
investigation of polynomial superalgebras is discussed, in relation to symmetries of
classical and quantum systems, including detailed comparisons with
previous studies in the literature.  Applications of the present work
include supersymmetry between colour singlet baryon and meson states
in Hamiltonian lattice QCD, and (for $n=4$) new classes of conformal
spacetime supersymmetries. The appendix, \S A provides
notational conventions for partition labelling of finite-dimensional
irreducible tensor representations of $gl(n)$ (\S A.1), and generalised Gel'fand
notation for the generators (\S A.2). In \S A.3 details of the fermionic oscillator
construction for the case $gl_{2}(n/ \{ 3\} + \{ \overline{3} \} )$ 
are given,
allowing (indecomposable) modules to be identified in a $gl(n)$ basis,
both on the fermionic Fock space, and via the adjoint action on the
associated Clifford algebra (see tables \ref{tbl:OperatorsGl1} and
\ref{tbl:OperatorsGl2} for the $n=1$ and $n=2$ cases).  Finally (\S 
A.4), for the case $n=4$ the relation between the algebras
$gl_{2}(4/{\{}1^3{\}} + {\{}\overline{1^3}{\}})$ (discussed in \S 3)
and the family $gl_{2}(4/{\{}1{\}} +
{\{}\overline{1}{\}})^{\texttt{a},\alpha, \beta}$ (\S 4)
is studied.

\section*{\S 2 Polynomial super-$gl(n)$ algebras $gl_{k}(n/ \{ \lambda \} + 
\{ \overline{\lambda} \} )$}

In this section generic polynomial super-$gl(n)$ algebras will be 
studied, from the point of view of admissible structure constants in 
the generalised sense.  From the graded Jacobi identities (see \S 4 
below), the odd generators form an $L_{\overline{0}}$-module with 
respect to (the adjoint action of) the even subalgebra, and 
${\{}L_{\overline{1}}, L_{\overline{1}}{\}}$ transforms under 
$\mbox{ad}_{L_{\overline{0}}}$ in the (symmetric) tensor product 
$L_{\overline{1}} \otimes L_{\overline{1}}$ of the odd 
$L_{\overline{0}}$-module $L_{\overline{1}}$ with itself. 
Correspondingly, in view of the the Poincar\'{e}-Birkhoff-Witt 
theorem for the structure of the enveloping algebra, monomials in the even 
generators transform as symmetric tensor powers of the adjoint 
representation. Thus generalised structure constants\footnote{ If 
${\{} T_{a} {\} }$ is a basis for the even subalgebra 
$L_{\overline{0}}$, and ${\{} Q_{\alpha} {\} }$ a basis for 
$L_{\overline{1}}$, then the (anti)commutation relations 
(\ref{eq:GeneralDefL0L1}) take the form
\begin{eqnarray*}
{[}T_{a}, T_{b} {]} &=& {f_{ab}}^{c}T_{c}, \nonumber \\
{[}T_{a}, Q_{\beta} {]} &=& {f_{a\beta}}^{\gamma}Q_{\gamma}, \nonumber \\
{\{}Q_{\alpha}, Q_{\beta} {\}} &=& {f_{\alpha \beta}}^{c_{1} c_{2} 
\ldots}T_{c_{1}}T_{c_{2}}\ldots + \cdots, 
\end{eqnarray*}
where there may be lower degree terms in the last line. Thus ${\{} 
Q_{\alpha} {\} }$ form a tensor operator under the action of $L_{\overline{0}}$,
and the anticommutator ${\{}Q_{\alpha}, Q_{\beta} {\}}$ transforms in 
the tensor product of the relevant representations of the even 
subalgebra.} 
can only exist with the correct polynomial 
degree $k$ if the corresponding \textit{symmetric} $k$'th tensor power of the 
adjoint representation $\mbox{ad}_{ L_{\overline{0}} }$ contains common 
irreducible submodules, with the branching multiplicity of the latter determining 
their number and type.

Study of the classification of such superalgebras devolves to
examination of possible $L_{\overline{0}}$-modules $L_{\overline{1}}$,
and admissible structure constants (\ref{eq:GeneralDefL0L1}). Similar
questions arise in the study of simple Lie 
superalgebras\cite{ScheunertBook} (where
$L_{\overline{0}}$ replaces $ U(L_{\overline{0}})$ in
(\ref{eq:GeneralDefL0L1}) above).  An analogous situation is addressed 
in the Witt construction\cite{Witt}, where one considers Lie
algebras associated with a given Lie algebra $L_{\overline{0}}$
extended by a certain $L_{\overline{0}}$-module (a trivial extension
being the semi-direct product, with the module given the structure of
an abelian algebra)\footnote{This construction has recently been
considered in connection with embeddings of quantum 
algebras\cite{Hannabuss}.}.
These considerations are more tractable if we turn to the `type I'
super-$gl(n)$ algebras: the $L_{\overline{0}}$-module
$L_{\overline{1}}$ is the sum of a single irreducible representation
and its contragredient.  The ${\mathbb Z}_{2}$-grading is thus
inherited from a ${\mathbb Z}$-grading of $L$, associated with the
spectrum of the adjoint action of the abelian summand of $gl(n) \simeq
sl(n) + gl(1)$.  Thus $L = L_{-1}+ L_{0}+ L_{1}$, with
$L_{\overline{0}}= L_{0}$ and $L_{\overline{1}} = L_{-1}+ L_{1}$. 
This entails ${\{} L_{\pm 1}, L_{\pm 1}{\}} = 0$, and ${\{} L_{+1},
L_{-1}{\}} \subset U(L_{0})$.  Without loss of generality we may
assume that $L_{+1}$ is an irreducible representation of the
semisimple part $sl(n)$, with $L_{-1}$ the corresponding
contragredient.  Now for semisimple Lie algebras,
Joseph's theorem\cite{HumphreysBook} states that the enveloping algebra $U(L_{0})$ is
isomorphic as an $L_{0}$-module, to the sum over all dominant
integral weights, of the tensor product of the corresponding
(finite-dimensional) highest weight module, with its contragredient. 
Thus, it is possible to investigate whether the anticommutators ${\{}
L_{+1}, L_{-1}{\}}$ can be associated with a unique element of
$U(L_{0})$.  However, as Joseph's theorem does
not mandate any relation between the polynomial degree within
$U(L_{0})$ of elements of a given tensor product contributing to the
sum, we proceed more generally.  Let, then, ${\{} \lambda {\}}$ denote
a dominant integral weight of $gl(n)$ and ${\{} \overline{\lambda}
{\}}$ the corresponding contragredient; where no confusion arises,
these symbols will also stand for the character of the corresponding irreducible,
finite dimensional highest weight module (for notation see appendix,
\S A.1)\footnote{We label highest weight representations, and where 
no confusion arises their corresponding characters, by partitions
${\{} \lambda {\}} = {\{} \lambda_{1}, \lambda_{2}, \ldots,
\lambda_{n} {\}}$ corresponding to the symmetry type of irreducible
tensors.  Partition labelling for irreducible representations of
simple Lie algebras is developed in \cite{BlackKingWybourne83}. 
Aspects of this notation for $gl(n)$ and $sl(n)$, including the 
various `products' $\cdot$, $\otimes$, $\circ$, are discussed in the
appendix, \S A.1.  In particular, a composite partition ${\{}
\overline{\rho};\sigma {\}}$ corresponds to an irreducible tensor of
contravariant symmetry type $\rho$, and covariant symmetry type
$\sigma$, and traceless with respect to contractions between
contravariant and covariant indices.  See \S A.1, and
also \cite{King75}.}.  The adjoint
representation of $gl(n)$ is the reducible representation ${\{}
\overline{1} {\}}\cdot{\{} 1 {\}}$ (corresponding to the tensor 
product of the $n$-dimensional defining representation with its 
contragredient) with irreducible parts ${\{}
\overline{1};1 {\}}+ {\{} 0 {\}}$ reflecting to the reduction to
$sl(n) + gl(1)$.  According to the previous discussion, distinct types
of structure constant will be determined by the branching multiplicity
of those irreducible components of the symmetric $k$'th tensor power
of the adjoint representation, which are in common with the
irreducible modules occurring in the decomposition of the tensor
product ${\{} \overline{\lambda} {\}} \cdot {\{} \lambda {\}}$. 
Let $\ell$ be the weight of ${\{} \lambda {\}}$ as a partition, and
$k$ be the polynomial degree of nonlinearity in the enveloping algebra
of $gl(n)$ characterising the algebra.  Then we have,
\begin{eqnarray}
{\{} \overline{\lambda} {\}}\cdot {\{} \lambda {\}} &=& 
 \sum_{\mu , \nu} {n^{\lambda}}_{\mu \nu}{\{} \overline{\mu}; \nu {\}}; \nonumber \\
({\{} \overline{1} {\}}\cdot{\{} 1 {\}})\otimes {\{}k{\}} &=& 
  \sum_{\mu , \nu } {n^{k}}_{\mu \nu}{\{} \overline{\mu}; \nu {\}}.
\label{eq:TensorCouplingsResults}
\end{eqnarray}
The multiplicities ${n^{k}}_{\mu \nu}$, ${n^{\lambda}}_{\mu \nu}$ are
defined in \S A.1 in terms of the standard Littlewood-Richardson
coefficients.  In \S A.1 it is shown that $ {n^{k}}_{\mu
\nu} \ge {n^{\lambda}}_{\mu \nu}$ provided $k \ge \ell$, and moreover 
if $ \ell > k$, then ${n^{k}}_{\mu \nu}=0$ for $\mu, \nu \vdash k+1, \ldots,
\ell$. In the latter case, generalised structure constants
for the corresponding symmetry types ${\{} \overline{\mu}; \nu {\}}$
arising from ${\{} \overline{\lambda} {\}}\cdot {\{} \lambda {\}}$ do 
not exist.

To complete the discussion of couplings in specific cases, it is
necessary to adopt an explicit notation.  Recall the well-known
presentation of $gl(n)$ via the Gel'fand generators ${E^{a}}_{b},\, 1
\le a,b \le n$ (see for example the first line of
(\ref{eq:GlnGelfand}) below).  Define the matrix powers
${(E^{k+1})^{a}}\mbox{}_{b} = {(E^{k})^{a}}\mbox{}_{c}{E^{c}}_{b}$ in the obvious
way, and their traces $ \langle E^{k} \rangle = {(E^{k})^{c}}\mbox{}_{c}$
(the standard Casimir invariants, see \S A.2).  Bearing in mind
Joseph's theorem\cite{HumphreysBook}, 
the following generalised permanents\cite{LittlewoodBook} provide a 
suitable spanning set for $U(gl(n))$: for ${\{} \lambda {\}} \vdash 
\ell$, and ${\{} \mu 
{\}}$  an $\ell$-part partition (of weight $k$, with nonzero parts 
augmented by zeroes as necessary), define
\begin{equation}
	{{[} {\{} \lambda {\}};{\{} \mu 
	{\}}{]}^{a_{1}a_{2}\ldots a_{\ell}}}_{b_{1}b_{2}\ldots b_{\ell}} = 
	\frac{1}{\ell ! ^{2}}\sum_{\rho, \sigma \in S_{\ell}} \chi^{\lambda}(\rho 
	\sigma^{-1}) {(E^{\mu_{1}})^{a_{\sigma 1}}}_{b_{\rho 1}} 
	{(E^{\mu_{2}})^{a_{\sigma 2}}}_{b_{\rho 2}} \ldots 
	{(E^{\mu_{\ell}})^{a_{\sigma \ell}}}_{b_{\rho \ell}}.
	\label{eq:PermanentDefn}
\end{equation}
Here $\chi^{\lambda}$ is the irreducible character of $S_{\ell}$ 
corresponding to the class $\lambda$, and ${(E^{0})^{a}}\mbox{}_{b} 
\equiv {\delta^{a}}_{b}$.  Thus ${\{} \lambda {\}}$ determines the 
symmetry type ${\{} \overline{\lambda} {\}} \cdot {\{} \lambda{\}}$, 
and ${\{} \mu {\}}$ the distribution of tensor contractions, for 
polynomials in the generators ${E^{a}}_{b}$ belonging to the enveloping 
algebra $U(gl(n))$.  In this notation the matrix powers are of course 
${(E^{k})^{a}}_{b} \equiv {{[} {\{} 1{\}};{\{} k {\}}{]}^{a}}_{b}$, 
while the Casimir operators are simply found by contraction of any 
${[} {\{} \lambda {\}};{\{} \mu {\}}{]}$ with
\[
{{\Delta_{\lambda}}^{a_{1}a_{2}\ldots a_{\ell}}}_{b_{1}b_{2}\ldots b_{\ell}} =
	{{[} {\{} \lambda {\}};{\{} 0 
	{\}}{]}^{a_{1}a_{2}\ldots a_{\ell}}}_{b_{1}b_{2}\ldots b_{\ell}}
\]

For the cases considered in the sequel, 
elementary tensor notation suffices for explicit constructions. As an 
example let us analyse in detail the case $k = \ell  = 3$ and ${\{} \lambda 
{\}} = {\{} 2,1 {\}}$. Explicitly, we have (see \S A.1)
\begin{eqnarray}
{\{} \overline{2,\!1} {\}}\cdot {\{} 2,\!1 {\}} &=& {\{} \overline{2,\!1}; 2,\!1 {\}}+
{\{} \overline{2}; 2 {\}} + 2{\{} \overline{1}; 1 {\}}+{\{}0{\}}, 
\nonumber \\
({\{} \overline{1} {\}}\cdot{\{} 1 {\}})\otimes {\{}3{\}} &=& {\{} \overline{3}; 3 {\}}+ 
{\{} \overline{2,\!1}; 2,\!1 {\}}+{\{} \overline{1,\!1,\!1}; 1,\!1,\!1 {\}}+ 
2{\{} \overline{2}; 2 {\}}+ \mbox{} \nonumber \\
&&\mbox{} + 2{\{} \overline{1,\!1}; 1,\!1 {\}} +4{\{} 
\overline{1}; 1 {\}}+3{\{}0{\}},
\label{eq:MixedDecomp}
\end{eqnarray}
from which, at the level of reduced matrix elements, at degree three, 
the putative cubic algebra $gl_{3}(n/ {\{} 2,\!1 {\}} + {\{} 
\overline{2,\!1} {\}} )$ has 10 types of structure constant (or 9 
free parameters for the associated reduced matrix elements, up to 
overall normalisation), corresponding to the maximum multiplicities of 
the common irreducible components of the above two decompositions 
(excluding additional structure constants arising from lower degree).  
To complete the construction of couplings for this case, define the following 
objects in the enveloping algebra, 
\begin{eqnarray}
	{{\{}F \cdot F' \cdot F''{\}}^{abc}}_{pqr} & \equiv & ({F^{a}}_{p}{F'^{b}}_{q} + 
	{F^{b}}_{p}{F'^{a}}_{q}){F''^{c}}_{r} \nonumber \\
   && \mbox{}	- ({F^{c}}_{p}{F'^{b}}_{q} + {F^{a}}_{p}{F'^{c}}_{q}){F''^{a}}_{r} + \ldots ,
\label{eq:MixedSymmetryObjects}
\end{eqnarray}	
being of \textit{mixed} symmetry type ${[} {\{} 2,1{\}}$; ${\{} 
\mu_{1}, \mu_{2}, \mu_{3} {\}}{]}$ with respect to contravariant and covariant
indices (where `\ldots' represents three additional quartets of 
terms establishing mixed symmetry with respect to the $pqr$ 
label permutations (the terms shown explicitly possess mixed symmetry with 
respect to $abc$). $F$, $F'$, $F''$ stand for $E^{\mu_{1}}$, 
$E^{\mu_{2}}$, $E^{\mu_{3}}$ respectively with 
$\mu_{1}+\mu_{2}+\mu_{3}=3$). Such terms exist by (\ref{eq:TensorCouplingsResults}),
and the 10 couplings at cubic degree required by
(\ref{eq:MixedDecomp}) are schematically\footnote{ Traceless forms of
the objects (\ref{eq:MixedSymmetryObjects}) can of course be
constructed if required.}
\begin{eqnarray}
&{\{}E \cdot E \cdot E{\}}, \quad  {\{}E^{2} \cdot E\cdot \delta{\}}, \quad {\{}E \cdot 
E\cdot \delta{\}}\langle E \rangle; \nonumber \\ &{\{}E^{3} \cdot \delta \cdot \delta{\}}, 
\quad {\{}E^{2} \cdot \delta \cdot \delta{\}}\langle E \rangle , \quad
{\{}E \cdot \delta \cdot \delta{\}}\langle E^{2} \rangle, \quad 
{\{}E \cdot \delta \cdot \delta{\}}\langle E \rangle^{2}; \nonumber \\
& {\{}\delta \cdot \delta \cdot \delta{\}}\langle E^{3} \rangle, \quad
{\{}\delta \cdot \delta \cdot \delta{\}}\langle E^{2} \rangle \langle E
\rangle, \quad {\{}\delta \cdot \delta \cdot \delta{\}}\langle E 
\rangle^{3},
\label{eq:MixedSymmetryTerms}
\end{eqnarray}
where ${\{}\cdot \; \cdot \; \cdot{\}}$ indicates mixed permutation 
symmetry as in (\ref{eq:MixedSymmetryObjects}) above.

It is noteworthy that the `leading' irreducible component ${\{} 
\overline{2,\!1}; 2,\!1 {\}}$ has unit multiplicity in both expansions 
in (\ref{eq:MixedDecomp}) above, corresponding to a single reduced 
matrix element (which can be set to one by a choice of overall 
normalisation); from the appendix, \S A.1, it is apparent that 
${n^{\lambda}}_{\lambda \lambda}=1$ in general (so the same count of 
the `leading' component holds whenever $\ell = k$).  This circumstance 
is intimately related to a canonical construction (outlined in \S A.2, 
valid for all simple Lie algebras), in which the generators and 
defining relations of $gl(n)$ may be presented in terms of components 
${{\mathcal E}^{abc\ldots}}_{pqr\ldots}$ of an arbitrary tensor 
operator (of any rank and symmetry type), and the defining relations 
presented in a manner consistent with this.

\section*{\S 3 Low rank examples and oscillator constructions}

The general discussion of the previous section has identified a large
class of candidate nonlinear super-$gl(n)$ algebras, on the basis of
structure constants which are admissible on the grounds of $gl(n)$
invariance.  This guarantees the validity of the Jacobi identity
involving ${[} L_{\overline{0}}, {\{} L_{\overline{1}},
L_{\overline{1}} {\}} {]}$.  The remaining odd Jacobi identity
involving ${[} L_{\overline{1}}, {\{} L_{\overline{1}},
L_{\overline{1}} {\}} {]}$ can best be addressed in specific
low-dimensional cases, or via explicit constructions, to which we now 
turn.  

Consider for example the simplest case $n=1$, and the polynomial
superalgebra $gl_{k}(1/{\{} \ell {\}} + {\{} \overline{\ell} {\}} )$. 
The abelian algebra $gl(1)$ has a single generator $K$, and
one-dimensional representations (labelled as type ${\{} \ell
{\}}$)\footnote{Here $\ell$ is simply an additive charge quantum
number.}.  With odd generators denoted $\overline{Q}$ and $Q$ we have
the defining relations (taking $\ell =1$ without loss of generality)
\[
{[}K, Q{]} = -Q, \quad {[}K, \overline{Q}{]} = \overline{Q}, \quad  
{\{}\overline{Q}, Q {\}} = f(K)
\]
for some polynomial $f$ of degree $k$, together with ${\{} Q , Q {\}}$
$ = {\{} \overline{Q} , \overline{Q} {\}} =0$.  The graded Jacobi
identity entails
\[
{[} Q, {\{} \overline{Q}, Q {\}} {]} = {[} {\{}Q,  \overline{Q} {\}}, 
Q {]} - {[} \overline{Q}, {\{} Q , Q {\}} {]} = {[} {\{}Q,  \overline{Q} {\}}, 
Q {]},
\]
whereupon $f(K)$ is central, $f(K) \equiv H$. Thus 
we have trivially regained the structure of the Lie superalgebra 
$gl(1/1)$ (the algebra of supersymmetric quantum mechanics), with defining relations
\begin{eqnarray*}
	{[}K, Q{]} = -Q, &  & {[}K, \overline{Q}{]} = +\overline{Q},  \\
	{[}H, Q{]} = 0, &  & {[}H, \overline{Q}{]} = 0, \\
	{[}H, K{]} = 0, &\mbox{and} & {\{}\overline{Q}, Q {\}} = H. 
\end{eqnarray*}

For the remainder of this section we consider examples of polynomial
super-$gl(n)$ algebras via concrete oscillator realisations, in which
there is enough structure to evaluate the anticommutator of odd
generators explicitly.  This will verify, for these cases the general
analysis of \S 2 above, and at the same time guarantee all Jacobi
identities.  Thus, we take generating sets $a_{i}, {a^{i}}$,
$i=1,2,\ldots, n $, $b_{j}, {b^{j}}$, $j=1,2,\ldots, n $ of either
fermionic or bosonic creation and annihilation operators,
respectively, where $a^{i} \equiv {a_{i}}^{\dagger}$, $b^{i} \equiv
{b_{i}}^{\dagger}$, satisfying the canonical (anti)-commutation
relations
\begin{eqnarray}
{\{} a_{i}, {a^{j}} {\}} = {\delta_{i}}^{j}\one, && {\{} a_{i}, {a_{j}} {\}} ={\{} {a^{i}}, {a^{j}} {\}} =0, \\
{[} b_{i}, {b^{j}} {]} = {\delta_{i}}^{j}\one, && {[} b_{i}, {b_{j}} {]} = {[} b^{i}, {b^{j}} {]} = 0
\label{eq:ccarandccr}
\end{eqnarray}
respectively (below we also consider parastatistics realisations).
The even generators for $gl(n)$ are the usual quadratic combinations 
giving the Gel'fand basis,
\begin{equation}
{E^{i}}_{j} = {a^{i}} a_{j} + \mbox{const}\one, \; \mbox{or} \; {E^{i}}_{j} 
 = {b^{i}}b_{j} + \mbox{const}\one,
\label{eq:GelfandBF}
\end{equation}
with
\begin{eqnarray}
{[}{E^{i}}_{j}, {E^{k}}_{\ell} {]} &=& {\delta_{j}}^{k} {E^{i}}_{\ell} - 
{\delta^{i}}_{\ell}{E^{k}}_{j}, \nonumber \\
{[}{E^{i}}_{j}, {c^{k}} {]} &=& {\delta_{j}}^{k} {c^{i}}, \nonumber \\
{[}{E^{i}}_{j}, {c_{k}} {]} &=& -{\delta^{i}}_{k} {c_{j}},
\label{eq:GlnGelfand}
\end{eqnarray}
where $c \equiv a\; \mbox{\textit{or}}\; b$. The odd generators of the
nonlinear superalgebras will be composites in the bosonic and
fermionic oscillator modes, transforming in tensor representations of
various types.  In particular, monomials purely in fermionic or
bosonic creation operators are automatically antisymmetric or
symmetric in permutations of their mode labels due to their mutual
anticommutativity or commutativity, respectively (monomials in the
corresponding annihilation operators transform contragrediently).  It
is thus natural to consider such monomials as candidates for odd
generators of super-$gl(n)$ algebras.

To set the context for the generalisations under investigation, we
consider firstly the rank one and two cases, with either
${\{}\lambda {\}}= {\{} 1^{\ell} {\}}$ (fermions), or ${\{}\lambda
{\}}= {\{} {\ell} {\}}$ (bosons), $\ell = 1, 2$.  Surprisingly
perhaps, at this algebraic level, particle statistics does not
preclude utilisation of bosonic oscillators as odd generators, and it
will turn out that the $k=1$ cases are well known constructions (there
are also $k=2$ generalizations).  To make the discussion complete, we
also consider polynomial algebras with generators closing on 
commutation relations, and also polynomial 
superalgebras with even part \textit{larger} than $gl(n)$.

The rank $\ell =1$ case is familiar in the linear situation, and
corresponds simply to enlarging the bilinear oscillator $gl(n)$
generators by appending the mode operators themselves.  Take firstly
the bosonic generators under the `natural' commutator bracket
relations.  Clearly $b^{i}, b_{j}, \one$ and ${E^{i}}_{j}$ (see
(\ref{eq:ccarandccr}b) and (\ref{eq:GlnGelfand}, (\ref{eq:GelfandBF})
above) generate the standard semidirect product of the Weyl-Heisenberg
algebra with the automorphism algebra $gl(n)$ (extendible to $sp(2n)$,
see below).  In the context of \textit{nonlinear} constructs there is
also the remarkable Holstein-Primakoff-Dyson\cite{HolsteinPrimakoff}
realisation, which `dresses' the oscillator modes so that linear
closure on $gl(n+1)$ is obtained:
\[
{E^{n+1}}_{i} = \sqrt{p \one + \langle E \rangle} b_{i}, \quad 
{E^{i}}_{n+1} = b^{i} \sqrt{p \one  + \langle E \rangle}, 
\]
for some parameter $p$.

As is well known, the same bosonic modes are natural candidates for 
odd generators of a superalgebra. In this case, a 
super-$gl(n)$ algebra is not achieved, as the `unnatural' choice of 
\textit{anti}commutator will generate new operators $\overline{S}\mbox{}^{ij}\equiv \frac 
12{\{}b^{i}, b^{j}{\}}$, $S_{ij}\equiv \frac 12 {\{}b_{i}, b_{j}{\}}$, as 
well as ${E^{i}}_{j} \equiv \frac 12 {\{}b^{i}, b_{j}{\}}$ 
(see (\ref{eq:GelfandBF}) above). However, instead closure is achieved on 
the enlarged automorphism algebra $sp(2n)$ in the `even' (bilinear) 
generators, and hence on the natural superalgebra $osp(1/2n)$  
including the oscillator modes themselves.

The situation with fermionic oscillators is entirely parallel to the 
bosonic case.  Choosing closure of the oscillator modes by 
anticommutators reproduces the canonical anticommutation relations, 
giving the complex Dirac or Clifford algebra generated by $a^{i}, 
a_{j}, \one$ together with automorphisms generated by ${E^{i}}_{j}$ 
(extendible to $so(2n)$, see below).  Again, there is a construction 
analogous to the Holstein-Primakoff-Dyson realisation 
\cite{DondiJarvis80,DondiJarvis81,GouldBrackenHughes,Palev96}, this 
time formally polynomial rather than in an extension of the enveloping 
algebra, whereby the fermionic generators can be appropriately 
`dressed' so as to achieve closure, this time on the Lie superalgebra 
$sl(n/1)$:
\[
{E^{n+1}}_{i} = \sqrt{p \one - \langle E \rangle} a_{i}, \quad 
{E^{i}}_{n+1} = a^{i} \sqrt{p \one - \langle E \rangle}
\]
for some parameter $p$.
Finally the choice of `unnatural' \textit{commutator} brackets for the
fermionic modes will generate, as well as ${E^{i}}_{j} \equiv \frac 
12{[}a^{i}, a_{j}{]}$ (see (\ref{eq:GelfandBF}) above), new operators
$\overline{A}\mbox{}^{ij}\equiv \frac 12{[}a^{i}, a^{j}{]}$, $A_{ij}\equiv \frac 12 
{[}a_{i}, a_{j}{]}$ which together with $a^{i}$, $a_{j}$ close on
the enlarged automorphism algebra $so(2n+1)$.  For the choice of
\textit{commutation} relations, the rank $\ell=2$ case is subsumed in
the above discussion of $\ell=1$, in that closure on $sp(2n) \supset
gl(n)$, $so(2n) \supset gl(n)$ was already found for bosons and
fermions respectively via the symmetric and antisymmetric rank two
tensors $S$ and $A$.

With the exception of the Holstein-Primakoff-Dyson realisation and its 
superalgebra analogue, all examples so far have been for up to 
quadratic realisations of classical Lie (super) algebras ($k=1$).  
This subject can be refined to deal with many cases of subalgebra 
chains, and especially to discuss real forms\cite{Nowicki,Gunaydin}.  
On the other hand, the behaviour of the $S$ and $A$ tensors under 
\textit{anti}commutation with their contragredients furnishes a first 
example of polynomial superalgebras, although not of ${\mathbb Z}_{2}$ 
graded super-$gl(n)$ type:
\begin{eqnarray}
{\{} \overline{S}\mbox{}^{(ij)}, S_{(pq)} {\}} &=& \frac 12 {(E \cdot E)^{(ij)}}_{(pq)} + \frac 
12 {(E \cdot \delta)^{(ij)}}_{(pq)} + {\delta^{(ij)}}_{(pq)}, 
\nonumber \\
{\{} \overline{A}\mbox{}^{[ij]}, A_{[pq]} {\}} &=& - \frac 12 {[E \cdot 
E]^{[ij]}}_{[pq]} + \frac 
12 {[E \cdot \delta]^{[ij]}}_{[pq]} - {\delta^{[ij]}}_{[pq]},
\label{eq:UnnaturalAnticommutators}	
\end{eqnarray}
where $(E \cdot E)$, $(E \cdot \delta)$, ${[}E \cdot E{]}$, ${[}E \cdot 
\delta {]}$ are the strength one minimal combinations\footnote{For simplicity 
the $gl(n)$ generators are defined as ${E^{i}}_{j} \equiv b^{i}b_{j}$, 
${E^{i}}_{j} \equiv a^{i}a_{j}$ in these equations (see 
(\ref{eq:GelfandBF}) above).} possessing the 
appropriate (anti)symmetry (compare (\ref{eq:PermanentDefn}) above),
\begin{eqnarray*}
{{(}E \cdot E {)}^{(ij)}}_{(pq)} &=& 
	({E^{i}}_{p}{E^{j}}_{q} + {E^{j}}_{p}{E^{i}}_{q} + 
	           {E^{i}}_{q}{E^{j}}_{p} + {E^{j}}_{q}{E^{i}}_{p}), \\	
{{(}E \cdot \delta {)}^{(ij)}}_{(pq)} &=& 
	({E^{i}}_{p}{\delta^{j}}_{q} + {E^{j}}_{p}{\delta^{i}}_{q} + 
	           {E^{i}}_{q}{\delta^{j}}_{p} + {E^{j}}_{q}{\delta^{i}}_{p}), \\
{\delta^{(ij)}}_{(pq)} &=& {\delta^{i}}_{p}{\delta^{j}}_{q} + 
{\delta^{i}}_{q} {\delta^{j}}_{p}, \\			   
			   {{[}E \cdot E {]}^{[ij]}}_{[pq]} &=& 
	({E^{i}}_{p}{E^{j}}_{q} - {E^{j}}_{p}{E^{i}}_{q} - 
	           {E^{i}}_{q}{E^{j}}_{p} + {E^{j}}_{q}{E^{i}}_{p}), \\
{{[}E \cdot \delta {]}^{[ij]}}_{[pq]} &=& 
	({E^{i}}_{p}{\delta^{j}}_{q} - {E^{j}}_{p}{\delta^{i}}_{q} - 
	           {E^{i}}_{q}{\delta^{j}}_{p} + {E^{j}}_{q}{\delta^{i}}_{p}), \\
{\delta^{[ij]}}_{[pq]} &=& {\delta^{i}}_{p}{\delta^{j}}_{q} - 
{\delta^{i}}_{q} {\delta^{j}}_{p}.
\end{eqnarray*}	
Including also the mutual \textit{commutativity} of the tensor components of 
each of these tensors separately, gives a structure of mixed grading resembling a 
nonlinear colour (super)algebra\footnote{
The natural graded structure when both fermionic and bosonic 
oscillator modes are present, where closure on \textit{both} linear and bilinear 
combinations is required, is indeed that of a ${\mathbb Z}_{2}\times
{\mathbb Z}_{2}$ graded colour superalgebra. See 
\cite{JarvisWybourneYangMei} and references therein.}. 

For the remaining two concrete examples in this section, we move to
the rank $\ell =3$ case, with fermionic oscillator realisations for
either ${\{} \lambda {\}} = {\{} 1^{3} {\}}$, $k=2$ or ${\{} \lambda
{\}} = {\{}{3} {\}}$, $k=2$.  These two cases typify several infinite
families of super-$gl(n)$ algebras with analogous structure (in which
$k$ grows with $\ell$).  Further examples, including ${\{} \lambda
{\}} = {\{} 2,1 {\}}$ at rank 3 in a parafermionic construction, and
higher-dimensional bosonic cases, are discussed in more general terms
in the conclusion of this section.

For $gl_{2}(n/{\{} 1^{3} {\}} + {\{} \overline{1^{3}} {\}} )$ define 
the odd generators
\begin{equation}
\overline{Q}\mbox{}^{[ijk]} =  a^{i}a^{j}a^{k}, \quad Q_{[pqr]} = 
a_{p}a_{q}a_{r},  
\label{eq:QbarQDefns}
\end{equation}
together with 
\begin{equation}
	{E^{i}}_{j} \equiv a^{i}a_{j}
	\label{eq:EijDefn}
\end{equation}
(see (\ref{eq:GelfandBF})). Then, in 
appropriately symmetrised tensor notation, the defining relations of 
this realisation of $gl_{2}(n/{\{} 1^{3} {\}} + {\{} \overline{1^{3}} {\}} 
)$ become\footnote{
Following the arguments of \S 2 above, in this case 
there are potentially 8 couplings, or 7 arbitrary coefficients up to 
normalisation (see \S A.1 and (\ref{eq:AdjointPlethysm}) below).}
\begin{eqnarray}
	{\{} \overline{Q}\mbox{}^{[ijk]}, Q_{[pqr]} {\}} &=&
	-\textstyle{\frac 14} {{[}E \cdot E \cdot \delta{]}^{[ijk]}}_{[pqr]} 
	+\textstyle{\frac 12}{{[}E \cdot 
	\delta \cdot \delta{]}^{[ijk]}}_{[pqr]} - 
	{\delta^{[ijk]}}_{[pqr]} \nonumber \\
	{\{} Q^{[ijk]}, Q^{[pqr]} {\}} &=& 0, \quad {\{} Q_{[ijk]}, Q_{[pqr]} 
	{\}} = 0, \nonumber \\
	{[}{E^{i}}_{j}, \overline{Q}\mbox{}^{[k \ell m]}{]} &=& 
	{\delta_{j}}^{k}\overline{Q}\mbox{}^{[i \ell m]} + 
	{\delta_{j}}^{\ell}\overline{Q}\mbox{}^{[k i m]} + 
	{\delta_{j}}^{m}\overline{Q}\mbox{}^{[k \ell i]}, \nonumber \\
	{[}{E^{i}}_{j}, {Q}_{[pqr]}{]} &=& 
	-{\delta^{i}}_{p} {Q}_{[jqr]} -{\delta^{i}}_{q}{Q}_{[pjr]} 
    -{\delta^{i}}_{r}{Q}_{[pqj]}.	
		\label{eq:SuperGl1^3}
\end{eqnarray}
Here  $[E \cdot E \cdot \delta]$ and $[E \cdot \delta \cdot \delta]$ 
are the appropriate strength one minimal combinations possessing the 
required \textit{anti}-symmetry (compare (\ref{eq:PermanentDefn}) above),
\begin{eqnarray}
	{{[}E \cdot E \cdot \delta{]}^{[ijk]}}_{[pqr]} &=& 
	({E^{i}}_{p}{E^{j}}_{q} - {E^{j}}_{p}{E^{i}}_{q} - 
	           {E^{i}}_{q}{E^{j}}_{p} + {E^{j}}_{q}{E^{i}}_{p}){\delta^{k}}_{r} + \ldots , \nonumber \\
    {{[}E \cdot \delta \cdot \delta{]}^{[ijk]}}_{[pqr]} &=& 
	{E^{i}}_{p}({\delta^{j}}_{q}{\delta^{k}}_{r} - {\delta^{j}}_{r}{\delta^{k}}_{q}) + \ldots , \nonumber \\
    {\delta^{[ijk]}}_{[pqr]} &=& {\delta^{i}}_{p} ({\delta^{j}}_{q} {\delta^{k}}_{r} - {\delta^{j}}_{r}{\delta^{k}}_{q})+ \ldots.
   \label{eq:AntiSymmetricObjects}
\end{eqnarray}
Allowing for cyclic permutations on $ijk$ and $pqr$, ${[}E \cdot E \cdot 
\delta{]}$ contains a total of $4\times 9$ 
$=36$ terms, ${[}E \cdot \delta \cdot \delta{]}$ contains $9 \times 
2 = 18$ terms, and ${[}\delta \cdot \delta \cdot \delta{]}$	just $3 
\times 2 = 6$ terms.
	
For $gl_{2}(n/{\{}{3} {\}} + {\{} \overline{{3}} {\}} )$ introduce 
$m = 3n$ and corresponding fermionic oscillators $a^{iA}, a_{jB}$, 
$i,j = 1,\ldots, n$ and $A,B = 1,2,3$. We take (see 
(\ref{eq:GelfandBF}))
\begin{equation}
{E^{iA}}_{jB} \equiv {a^{iA}} a_{jB}, \quad 
{E^{i}}_{j} = {E^{iA}}_{jA}, \quad {F^{A}}_{B} = {E^{iA}}_{iB},
\label{eq:ColourEijDefn}
\end{equation}
for the generators of $gl(3n)$, $gl(n)$ and colour $gl(3)$, 
respectively. Then with the help of the 
totally antisymmetric alternating tensor $\epsilon_{ABC}$ define
\begin{equation}
\overline{W}\mbox{}^{(ijk)} =  \epsilon_{ABC}a^{iA}a^{jB}a^{kC}, \quad
W_{(pqr)} =  \epsilon^{ABC}a_{pA}a_{qB}a_{rC}.
\label{eq:WbarWDefns}
\end{equation}
Clearly, the `colour singlet' combinations $\overline{W}\mbox{}^{(ijk)}$ and 
$W_{(pqr)}$ 
are \textit{symmetric} in the mode labels, and elementary calculation 
leads to the following, appropriately symmetrised, 
$gl_{2}(n/{\{}{3} {\}} + {\{} \overline{{3}} {\}} )$ defining 
relations\footnote{
Following the arguments of \S 2 above, in this case 
there are again potentially 8 couplings, or 7 arbitrary coefficients up to 
normalisation (see \S A.1 and (\ref{eq:AdjointPlethysm}) below).}:
\begin{eqnarray}
	{\{} W^{(ijk)}, W_{(pqr)} {\}} &=&
	\frac 12{(E \cdot E \cdot \delta)^{(ijk)}}_{(pqr)} +2{(E \cdot 
	\delta \cdot \delta)^{(ijk)}}_{(pqr)} 
	-6{\delta^{(ijk)}}_{(pqr)} \nonumber \\
	{\{} W^{(ijk)}, W^{(pqr)} {\}} &=& 0, \quad {\{} W_{(ijk)}, W_{(pqr)} 
	{\}} = 0, \nonumber \\
	{[}{E^{i}}_{j}, \overline{W}\mbox{}^{(k \ell m)}{]} &=& 
	{\delta_{j}}^{k}\overline{W}\mbox{}^{(i \ell m)} + 
	{\delta_{j}}^{\ell}\overline{W}\mbox{}^{(k i m)} + 
	{\delta_{j}}^{m}\overline{W}\mbox{}^{(k \ell i)}, \nonumber \\
	{[}{E^{i}}_{j}, {W}_{(pqr)}{]} &=& 
	-{\delta^{i}}_{p} {W}_{(jqr)} -{\delta^{i}}_{q}{W}_{(pjr)} 
    -{\delta^{i}}_{r}{W}_{(pqj)}.	
		\label{eq:SuperGl3}
\end{eqnarray}
Here  $(E \cdot E \cdot \delta)$ and $(E \cdot \delta \cdot \delta)$ 
are the appropriate strength one minimal combinations possessing the 
required \textit{symmetry} type (compare (\ref{eq:PermanentDefn}), 
(\ref {eq:AntiSymmetricObjects}) above),
\begin{eqnarray}
	{{(}E \cdot E \cdot \delta{)}^{(ijk)}}_{(pqr)} &=& 
	({E^{i}}_{p}{E^{j}}_{q} + {E^{j}}_{p}{E^{i}}_{q} + 
	           {E^{i}}_{q}{E^{j}}_{p} + {E^{j}}_{q}{E^{i}}_{p}){\delta^{k}}_{r} + \ldots , \nonumber \\
    {{(}E \cdot \delta \cdot \delta{)}^{(ijk)}}_{(pqr)} &=& 
	{E^{i}}_{p}({\delta^{j}}_{q}{\delta^{k}}_{r} + {\delta^{j}}_{r}{\delta^{k}}_{q}) + \ldots , \nonumber \\
    {\delta^{(ijk)}}_{(pqr)} &=& {\delta^{i}}_{p} ({\delta^{j}}_{q} 
    {\delta^{k}}_{r} + {\delta^{j}}_{r}{\delta^{k}}_{q})+ \ldots.
   \label{eq:SymmetricObjects}
\end{eqnarray}
Allowing for cyclic permutations on $ijk$ and $pqr$, $(E \cdot E \cdot
\delta)$ contains a total of $4\times 9$ $=36$ terms, $(E \cdot \delta
\cdot \delta)$ contains $9 \times 2 = 18$ terms, and $(\delta \cdot
\delta \cdot \delta)$ just $3 \times 2 = 6$ terms.

We close this section with some general remarks on ways of furnishing
further specific constructions of polynomial superalgebras.  It is
clear that the two rank three fermionic examples generalise to
arbitrary rank.  In the antisymmetric ${\{} \lambda {\}} = {\{}
1^{\ell} {\}}$ case, choose degree $\ell$ monomials in the fermionic
creation and annihilation operators (giving anticommutators closing at
degree $k = \ell -1$ in ${E^{i}}_{j}$), and in the symmetric ${\{}
\lambda {\}} = {\{} {\ell} {\}}$ case, choose $m=\ell n$ and define
(for odd degree $\ell$) colour singlet monomials with the appropriate
rank $\ell$ alternating tensor, with anticommutators closing on degree
$k= \ell-1$ in ${E^{i}}_{j}$.

More generally, contractions with \textit{any} suitable invariant
tensor yield a plethora of possible symmetry types ${\{} \lambda {\}}$
for putative odd generators.  For example, if $m = pn$ and a symmetric
bilinear form (metric tensor) of dimension $p$ exists, it is known
that tensor powers yield symmetry types corresponding to all
partitions ${\{} \lambda' {\}}$ of even row lengths (see \S A.1).  The
corresponding tensor contraction against a totally antisymmetric
monomial
\[
a^{i_{1}A_{1}}\ldots a^{i_{\ell}A_{\ell}}
\]
thus yields a fermionic tensor of $gl(n)$ symmetry type ${\{} \lambda
{\}}$ (with even column lengths) corresponding to the transpose of the
partition ${\{} \lambda' {\}}$ (see \S A.1).  Similar remarks apply to
tensors constructed from an available \textit{antisymmetric} bilinear
form (with $p$ even).  However, as $\ell$ is necessarily even, closure
with \textit{anti}commutators is `unnatural', and the candidate
polynomial superalgebra is of $so(2n)$ type in this case.  In
principle, similar purely bosonic constructions are available.  From
the discussion of the rank one and two cases, at rank 3 in the
symmetric case, a polynomial super-$sp(2n)$ algebra of the type
$sp_{3}(2n/ {\langle}3 {\rangle})$ can be expected, with corresponding
higher-dimensional generalisations.  A more general approach, but
beyond the scope of the present work, is to appeal to the
well-developed theory of general tensor invariants of arbitrary rank
and symmetry type\cite{MacFarlane,Okubo,King,Molev}.

Despite the above flexibility and ubiquity in oscillator 
constructions, it is nonetheless difficult to see how a natural 
candidate polynomial superalgebra for \textit{odd} rank \textit{mixed} 
symmetry tensor types is possible (few algebras have 
natural primitive invariants of this type).  A final possibility worth 
pointing out is the use of parafermi or parabose oscillator 
realisations\cite{Green53,Greenberg64}\footnote{Modular statistics 
also provide representations of colour algebras and superalgebras of 
general permutation symmetry type\cite{JarvisGreen83}.}. For example, 
for parafermions of order $p$, monomials of permutation symmetry type 
of up to $p$ columns in the mode labels can be constructed; indeed the 
fundamental trilinear relation
\[
{[}a^{i},{[}a^{j}, a^{k}{]}{]} = 0
\]
(for any order of parafermi statistics) simply ensures that the
combination $a^{i}{[}a^{j}, a^{k}{]}$ is automatically of 
mixed\cite{Hartle}
symmetry type ${\{}2,1{\}}$.  Furthermore, the structure of the
parafermionic oscillator enveloping algebra is such\cite{BrackenGreenNC} that even
monomials can be represented as polynomials in the bilinears
${[}a^{i}, a^{j}{]}$, ${[}a^{k}, a_{\ell}{]}$, ${[}a_{m}, a_{n}{]}$,
which are known to generate $so(2n)$ just as in the fermionic 
case\cite{BrackenGreenNC}. 
Thus the mixed symmetry rank three case realised by parafermions may
be a candidate $so_{3}(2n/{[}2,1{]})$ polynomial 
superalgebra\footnote{Partitions labelling irreducible 
representations of the symplectic (see above) and orthogonal Lie algebras are 
denoted $\langle \lambda \rangle$, ${[}\lambda {]}$ respectively. 
See \S A.1 and \cite{BlackKingWybourne83}.}. For
parafermi statistics of order $p$, the corresponding generalised Fock
space realisation\cite{BrackenGreenNC} would form a submodule of the spinor 
representation
${[} \textstyle{\frac 12}p,\textstyle{\frac 12}p \ldots, \pm
\textstyle{\frac 12}p{]}$ of $so(2n)$.

\section*{\S 4 Quadratic super-$gl(n)$ algebras $gl_{2}(n/ {\{} 1 {\}} +
{\{} \overline{1} {\}} )$}

In this section we develop a more
complete treatment of a single class of polynomial super-$gl(n)$
algebras than has been possible for the more general cases.  We return
to analogues of the simple Lie superalgebra $sl(n/1)$, wherein the
even part $gl(n) \simeq sl(n) + gl(1)$ is graded by odd generators in
the irreducible $n$-dimensional \textit{defining} representation and
its contragredient.  In the present notation, we have $sl(n/1) \equiv
gl_{1}(n/ {\{} 1 {\}} + {\{} \overline{1} {\}} )$.  As shown in \S 3
above, the linear case is familiar from elementary oscillator
constructions, but we concentrate here on the quadratic
generalisation, $gl_{2}(n/ {\{} 1 {\}} + {\{} \overline{1} {\}} )$,
which cannot be so realised (except for the special case $n=4$, see
below).  Below, a complete account is given of structure constants and
defining relations, followed by a discussion of certain classes of
irreducible representations of these quadratic superalgebras.

Following the previous discussion of classes of structure constants, 
in order to write the nonvanishing anticommutator of the odd 
generators in the most general way, allowing for terms of degree 0, 1 
and 2 in the $gl(n)$ enveloping algebra, the following decompositions 
should be noted (see (\ref{eq:MixedDecomp}) and \S 2):
\begin{eqnarray}
{\{} \overline{1}{\}}\cdot {\{} {1}{\}} &=& {\{} \overline{1}; 1 
{\}}+{\{}0 {\}}, \nonumber \\
{\{} \overline{1}; 1 {\}} \otimes {\{} 2 {\}}+ {\{} \overline{1}; 1 {\}} 
\otimes {\{} 1 {\}}+
{\{} \overline{1}; 1 {\}} \otimes {\{} 0 {\}}
&=& ({\{} \overline{2}; 2 {\}}+{\{} \overline{1,1}; 1,1 {\}}+ 2{\{} \overline{1}; 1 {\}}
+2 {\{} 0{\}})+\mbox{} \nonumber \\
&& \mbox{} + ({\{} \overline{1}; 1 {\}}+{\{}0{\}}) + ({\{}0{\}})
\label{eq:AdjointPlethysm}
\end{eqnarray}
so that there are seven couplings (or six arbitrary coefficients up to 
an overall normalisation). 
Independent terms are most conveniently expressed in an $sl(n)+gl(1)$ 
basis, for which we introduce the generators ${J^{a}}_{b}$, 
$\widehat{N}$, $\overline{Q}\mbox{}^{a}$, $Q_{b}$, $1 \le a,b,c \le n$ 
defined as follows:
\begin{eqnarray}
	{J^{a}}_{b} &\equiv& {E^{a}}_{b} - {\frac{1}{n}} {\delta^{a}}_{b}\widehat{N}, \quad 
	\widehat{N} \equiv \langle E \rangle = {E^{c}}_{c},  
	\nonumber \\
	{[} {J^{a}}_{b}, {J^{c}}_{d} {]} &=& 
	{\delta_{b}}^{c}  {J^{a}}_{b} -  {\delta^{a}}_{d} {J_{c}}^{b}, \quad 
	{[} \widehat{N}, {J^{a}}_{b} {]} = 0, \nonumber\\
	{[} {J^{a}}_{b}, \overline{Q}\mbox{}^{c} {]} &=& 
	{\delta_{b}}^{c}\overline{Q}\mbox{}^{a} - { \frac{1}{n}} 
	{\delta^{a}}_{b}\overline{Q}\mbox{}^{c}, \quad {[} \widehat{N}, 
	\overline{Q}\mbox{}^{c} {]} = \overline{Q}\mbox{}^{c}, \nonumber \\
	{[} {J^{a}}_{b},{Q}_{c} {]} &=& - {\delta^{a}}_{c} Q_{b} + {\frac{1}{n}} 
	{\delta^{a}}_{b} Q_{c}, \quad {[} \widehat{N}, 
	Q_{c}{]} = - Q_{c}.
\label{eq:DetailedCR}
\end{eqnarray}
Finally the general anticommutator is written in terms of six 
arbitrary coefficients,
\begin{eqnarray}
	{\{}\overline{Q}\mbox{}^{a}, Q_{b}{\}} &=& {(J^{2})^{a}}_{b} + 
	\texttt{a} \langle 
	J^{2}\rangle {\delta^{a}}_{b} + (\texttt{b}_{1} \widehat{N} + \texttt{b}_{2}) {J^{a}}_{b}
	+ (\texttt{c}_{1}\widehat{N}^{2} + \texttt{c}_{2}\widehat{N} + \texttt{c}) 
	{\delta^{a}}_{b}, \nonumber \\
	&\mbox{with}& \quad {\{} \overline{Q}\mbox{}^{a}, \overline{Q}\mbox{}^{b} {\}} = 
	0 = {\{} Q_{a}, Q_{b} {\}}. 
\label{eq:AntiCommutatorQbarQ}
\end{eqnarray}
The coefficients $\texttt{a}$, $\texttt{b}_{1}, \texttt{b}_{2}$, $\texttt{c}_{1}, \texttt{c}_{2}, \texttt{c}$ are 
determined by demanding that (\ref{eq:DetailedCR}), 
(\ref{eq:AntiCommutatorQbarQ}) above are consistent with the Jacobi 
identity,
\begin{equation}
{[}x,{[}y,z{]} = {[}{[}x,y{]},z{]} + (-1)^{(x)\cdot(y)} {[}y, {[}x,z{]}{]},
\end{equation}
for homogeneous $x$, $y$, $z$ $\in L$ with $(x), (y) = \overline{0} 
\mbox{ or } \overline{1}$ being the ${\mathbb Z}_{2}$-grading of $x, 
y$, respectively. In view of the cyclic symmetry, there are four 
choices of three homogeneous elements, namely 
$\overline{0}\overline{0}\overline{0}$, $\overline{0}\overline{0}\overline{1}$,
$\overline{0}\overline{1}\overline{1}$, $\overline{1}\overline{1}\overline{1}$,
of which the first is simply the Jacobi identity for 
$L_{\overline{0}}$, while the second and third express the covariance of 
$L_{\overline{1}}$ and ${\{} L_{\overline{1}}, L_{\overline{1}} {\}}$ 
under $\mbox{ad}_{L_{\overline{0}}}$ (which has been already built 
into (\ref{eq:AntiCommutatorQbarQ})). By similar reasoning, the only nontrivial Jacobi 
identities involving three odd elements are
\begin{eqnarray}
{[}{\{} \overline{Q}\mbox{}^{a}, Q_{b}{\}}, Q_{c}{]} &=& 
{[}\overline{Q}\mbox{}^{a}, {\{}Q_{b}, Q_{c} {\}} {]} - {[}{\{} 
\overline{Q}\mbox{}^{a}, Q_{c}{\}}, Q_{b}{]} \equiv - {[}{\{} 
\overline{Q}\mbox{}^{a}, Q_{c}{\}}, Q_{b}{]}, \nonumber \\
\mbox{and similarly} \quad {[}{\{} {Q}_{a}, \overline{Q}\mbox{}^{b}{\}}, \overline{Q}\mbox{}^{c}{]} &=&
- {[}{\{} {Q}_{a}, \overline{Q}\mbox{}^{c}{\}}, \overline{Q}\mbox{}^{b}{]}.
\label{eq:AntiSymmJacobiID}
\end{eqnarray}
Evaluating (\ref{eq:AntiSymmJacobiID}) explicitly using (\ref{eq:DetailedCR}), 
(\ref{eq:AntiCommutatorQbarQ}) above we have
\begin{eqnarray}
{[}{\{} \overline{Q}\mbox{}^{a}, Q_{b}{\}}, Q_{c}{]} &=&{[} 
- (Q \cdot J)_{b}{\delta^{a}}_{c} - 2\texttt{a} (Q \cdot J)_{c}{\delta^{a}}_{b}{]}+ 
{[}- Q_{b}{J^{a}}_{c} \!+\!(\texttt{b}_{1}\!-\!\frac{2}{n} ) Q_{b}{J^{a}}_{c}{]} + \mbox{} \nonumber \\
&& {[}(-\texttt{b}_{1}) Q_{b}\widehat{N}{\delta^{a}}_{c}+ 
(2\texttt{c}_{1}\!-\!\frac{\texttt{b}_{1}}{n}) Q_{c}\widehat{N}{\delta^{a}}_{b}{]} 
+\mbox{} \nonumber \\
&&{[}-\!(\texttt{b}_{2}\!-\!\frac{1}{n}) Q_{b}{\delta^{a}}_{c}+ 
(\frac{1}{n^{2}} - \frac{(n^{2}\!-\!1)}{n}\texttt{a}\!-\! 
   \frac{1}{n}\texttt{b}_{2}\!-\! 
   \texttt{c}_{1}\!+\!\texttt{c}_{2})Q_{c}{\delta^{a}}_{b}{]}, \nonumber
\end{eqnarray}
where $(Q \cdot J)_{a} \equiv Q_{b}{J^{b}}_{a}$. Similarly
\begin{eqnarray}
	{[}{\{} \overline{Q}\mbox{}^{a}, Q_{b}{\}}, \overline{Q}\mbox{}^{c}{]} & = & 
    {[} (J \cdot \overline{Q})^{a}{\delta^{c}}_{b} \!+\! 
2\texttt{a} (J \cdot \overline{Q})^{c}{\delta^{a}}_{c} {]} 
    + {[}\overline{Q}\mbox{}^{a}{J^{c}}_{b} \!-\!(\texttt{b}_{1}\!-\!\frac{2}{n}) 
\overline{Q}\mbox{}^{c}{J^{a}}_{b} {]} + \mbox{} \nonumber \\
   & &	{[} \texttt{b}_{1} \widehat{N} \overline{Q}\mbox{}^{a}{\delta^{c}}_{b} 
   \!-\! (2\texttt{c}_{1}\!-\!\frac{\texttt{b}_{1}}{n})\widehat{N} 
   \overline{Q}\mbox{}^{c}{\delta^{a}}_{b} {]}+ \mbox{} \nonumber \\
   & & {[}(\texttt{b}_{2}\!-\!\frac{1}{n})\overline{Q}\mbox{}^{a}{\delta^{c}}_{b} 
   \!-\!(\frac{1}{n^{2}} - \frac{(n^{2}\!-\!1)}{n}\texttt{a}\!-\! 
   \frac{1}{n}\texttt{b}_{2}\!-\! \texttt{c}_{1}\!+\!\texttt{c}_{2})
   \overline{Q}\mbox{}^{c}{\delta^{a}}_{b}{]}, \nonumber
\end{eqnarray}
with $(J \cdot \overline{Q})^{a} \equiv {J^{a}}_{b}\overline{Q}\mbox{}^{b}$. 
Imposing (\ref{eq:AntiSymmJacobiID}) yields  
\begin{eqnarray}
	\texttt{a} = - \frac 12, &\quad & \texttt{b}_{1} = - \frac{n-2}{n}, \nonumber \\
	\texttt{c}_{1} &=& \frac{(n-1)(n-2)}{2n^{2}}, \nonumber \\
	\mbox{and} \quad \frac{n-1}{n}\texttt{b}_{2} + \texttt{c}_{2} &=& 
	-\frac{n-1}{2},
	\label{eq:RawCoeffs}
\end{eqnarray}
with \textit{no} restriction on $\texttt{c}$. Finally it is always possible to 
absorb $\texttt{b}_{2}$ by means of an appropriate shift 
$\widehat{N} \rightarrow \widehat{N}' \equiv \widehat{N} + 
\mbox{const}\one$, yielding the  
anticommutator for the quadratic algebra \textit{uniquely} determined 
up to the central term,	
\begin{eqnarray}
	{\{}\overline{Q}\mbox{}^{a}, Q_{b}{\}} &=& {(J^{2})^{a}}_{b} - {\frac 12} \langle 
	J^{2}\rangle {\delta^{a}}_{b} - \frac{n-2}{n} \widehat{N} 
	{J^{a}}_{b} \nonumber \\
 && \mbox{}+ {\delta^{a}}_{b} 
 {[}\displaystyle{\frac{(n-1)(n-2)}{2n^{2}}}\widehat{N}^{2} 
 -\frac{(n-1)}{2}\widehat{N}{]} + \texttt{c} {\delta^{a}}_{b}\one.
\label{eq:AntiCommutatorQbarQUnique}
\end{eqnarray}

From (\ref{eq:RawCoeffs}) it can be seen that 
a more flexible set of algebraic defining relations emerges if, in 
addition to the anticommutation relations, covariant quadratic identities 
between the even and odd generators exist in the enveloping 
algebra. It is evident in any case from \S 2 and the 
examples of \S 3, that additional Serre-type relations can be expected 
for the consistency of the nonlinear algebras in general, and the present 
discussion is a case in point. Thus we impose the covariant 
conditions\footnote{Further details of the generalisation of these 
Serre-type relations to the cases $gl_{2}(n/{\{} 1^{3} {\}} + {\{} \overline{1^{3}} 
{\}} )$ and $gl_{2}(n/{\{}{3} {\}} + {\{} \overline{{3}} {\}} )$, 
as well as the present case, are given in \S A.4.}
\begin{eqnarray}
	(J \cdot \overline{Q})^{a} & = & (\overline{\alpha} \widehat{N} + 
	\overline{\beta}\one) \overline{Q}\mbox{}^{a}, \nonumber \\
	(Q \cdot J)_{a} & = & Q_{a}( \alpha \widehat{N} + \beta \one)
	\label{eq:ConstraintsJQbarQJ} 
\end{eqnarray}
for some constants $\alpha, \beta, \overline{\alpha}, 
\overline{\beta}$. This move releases the condition $\texttt{a}=-\frac 12$ 
found above; the remaining coefficients are determined as in 
(\ref{eq:RawCoeffs}), with modified constraints 
\begin{eqnarray}
	\texttt{b}_{1} &=& - \frac{n-2}{n}, \nonumber \\
\texttt{c}_{1} &=& \frac{(n-1)(n-2)}{2n^{2}} - (2\texttt{a}+1)\alpha, \nonumber \\
	\mbox{and} \quad \frac{n-1}{n}\texttt{b}_{2} + \texttt{c}_{2} &=& 
	-\frac{n-1}{2} - (2\texttt{a}+1)\beta,
	\label{eq:SerreCoeffs}
\end{eqnarray}
and a corresponding three parameter family $gl_{2}(n/ \{ 1 \} + 
\{ \overline{1} \} )^{\texttt{a},\alpha, \beta}$ of quadratic algebras.

Further remarks concerning the quadratic algebras, and 
the significance of the parameter $\alpha$ in constructing 
representations, are 
given in the conclusions, \S 5 below. Note that the 
quadratic antisymmetric rank 3 algebra, constructed via fermion 
annihilation and creation operators in \S 3, furnishes an example of 
this family, for the case $n=4$ (with the identification of the rank 
3 antisymmetric tensor representation with the contragredient of the 
defining representation for $sl(4)$). 
Define for $n=4$ $\overline{S}\mbox{}^{i} =\frac 16 
\epsilon^{ijk\ell}Q_{jk\ell}$, $S_{i} = \frac 16 \epsilon_{ijk\ell} 
\overline{Q}\mbox{}^{jk\ell}$, then from (\ref{eq:SuperGl1^3}) as 
shown in appendix \S A.4, 
\[
{\{} \overline{S}\mbox{}^{i}, S_{j}{\}} = {(J^{2})^{a}}_{b} -\frac 12 \langle 
	J^{2}\rangle {\delta^{a}}_{b} - \frac 12\widehat{N}{J^{a}}_{b}
	+ (\frac{3}{16}\widehat{N}^{2} -\frac 34 \widehat{N} + 2){\delta^{a}}_{b},
\]
which agrees with (\ref{eq:DetailedCR}), (\ref{eq:RawCoeffs}) for 
$n=4$ ($a=-\frac 12$, $\texttt{b}_{1}=-\frac 12$, $\texttt{c}_{1}= 
\frac {3}{16}$), together with
$\texttt{b}_{2}=-1, \texttt{c}_{2}= -\frac 34$. Details of the 
calculation, together with further consideration of the covariant 
identities (\ref{eq:ConstraintsJQbarQJ}), are provided in \S A.4.

As a final development we consider some aspects of the 
representation theory of the nonlinear super-$gl(n)$ algebras, as 
exemplified by the quadratic family $gl_{2}(n/ \{ 1 \} + \{ \overline{1} 
\} )^{\texttt{a},\alpha,\beta}$.  Parallels with the representation 
theory of classical superalgebras are brought out by the construction 
of induced modules, and consideration of their (a)typicality 
conditions.  Following Kac's construction for type I 
superalgebras\cite{Kac}, and in particular for $sl(n/1)$, an 
analogous induced module $\overline{V}{\{}\lambda{\}}$ for $gl_{2}(n/ \{ 1 
\} + \{ \overline{1} \} )^{\texttt{a},\alpha,\beta}$ may be constructed 
via the choice of a Borel superalgebra $B_{+}$, its associated 
enveloping superalgebra $U_{+}$, and an arbitrary\footnote{Note that 
here, in contrast to the previous notation, 
$\overline{V}{\{}\lambda{\}}$ is a Kac module based on an 
\textit{arbitrary} $L_{0}$-highest weight, but for the \textit{fixed} 
super-$gl(n)$ algebra $gl_{2}(n/ \{ 1 \} + \{ \overline{1} \} 
)$.} $L_{0}$-module 
$\overline{V}_{0}{\{}\lambda{\}}$ extended trivially to $B_{+}$.  Then 
with the help of the Poincar\'{e}-Birkoff-Witt theorem we have as 
usual
\[
\overline{V}{\{}\lambda{\}} \simeq \bigwedge(L_{-}) \otimes 
\overline{V}_{0}{\{}\lambda{\}},
\]
or explicitly, 
\[
\overline{V}{\{}\lambda{\}} = \sum_{k=0}^{n} \sum_{a_{1}, a_{2}, \ldots, 
a_{k}} Q_{a_{1}}Q_{a_{2}} \cdots Q_{a_{k}} \otimes 
\overline{V}_{0}{\{}\lambda{\}}.
\]
Introducing the highest weight vector $v^{+}$, of particular interest 
is the vector $Q_{n} \otimes v^{+}$, which (as it commutes with even 
raising operators ${E^{a}}_{b}$ for $1 \le a<b \le n$), will again be 
a be a $B_{+}$-highest weight vector, and moreover will cyclically 
generate an indecomposable submodule of $\overline{V}{\{}\lambda{\}}$ 
of highest weight \textit{different} from $\lambda$, if 
$\overline{Q}\mbox{}^{n} Q_{n} \otimes v^{+} =0$ (since 
$\overline{Q}\mbox{}^{a} Q_{n} \otimes_{U_{+}} v^{+} = {\{} 
\overline{Q}\mbox{}^{a}, Q_{n}{\}} v_{\lambda}^{+} =0$ for $a < n$).  
Thus a necessary condition for typicality is the nonvanishing of the 
eigenvalue of ${\{} \overline{Q}\mbox{}^{n}, Q_{n} {\}}$ on $v^{+}$.  
Similar considerations in fact apply to the hierarchy of vectors 
$Q_{n} \otimes v^{+}$, $Q_{n-1}Q_{n} \otimes v^{+}$, \ldots (see 
\cite{Kac}).

The connection with standard lexicographical (partition) labelling is 
simplest if the anticommutation relations 
(\ref{eq:AntiCommutatorQbarQUnique}) are re-written directly in terms 
of the standard $gl(n)$-generators ${E^{a}}_{b}$ (see (\ref{eq:DetailedCR})), 
yielding
\begin{eqnarray}
{\{}\overline{Q}\mbox{}^{a}, Q_{b}{\}} &\!=\!& {(E^{2})^{a}}_{b} -  \widehat{N} {E^{a}}_{b}
- \frac 12 {\delta^{a}}_{b}
{[} \langle E^{2}\rangle - \widehat{N}(\widehat{N}-n+1){]} + \texttt{c} {\delta^{a}}_{b}\one.
\label{eq:AntiCommutatorECase}
\end{eqnarray}
With the highest weight labels $\lambda_{1}, \lambda_{2},\ldots, 
\lambda_{n}$ of $gl(n)$ (eigenvalues of ${E^{1}}_{1}, {E^{2}}_{2}, 
\ldots, {E^{n}}_{n}$), for a dominant integral weight $\lambda$ such 
that $\lambda_{a}-\lambda_{b} \in {\mathbb Z}^{+}$ for $a > b$, and 
denoting the (eigenvalues of the) first and second degree Casimir operators  by $\Lambda$, 
$\texttt{C}$, we have for example for the eigenvalue $a_{n}$ of ${\{} 
\overline{Q}\mbox{}^{n}, Q_{n} {\}}$ by direct computation,
\begin{equation}
	a_{n}= \lambda_{n}(\lambda_{n}-\Lambda) -\frac 12 
	\texttt{C} +\frac 12 \Lambda(\Lambda-n+1)
	\label{eq:AtypicalityCondition}
\end{equation}
where the usual eigenvalues are understood:
\begin{equation}
\Lambda = \sum_{a=1}^{n} \lambda_{a}, \quad \texttt{C} = 
\sum_{a=1}^{n} \lambda_{a}(\lambda_{a} + n+1-2a).
\label{eq:CasimirEvalues}
\end{equation}

\section*{\S 5 Conclusions}

In this paper we have made a preliminary investigation of a large class 
of `polynomial super-$gl(n)$ algebras'. These mimic the classical 
simple Lie superalgebras $sl(n/1) \sim A(n-1,0)$ in possessing an 
even part $gl(n) \simeq sl(n) + gl(1)$, however with odd generators in 
an \textit{arbitrary} representation ${\{} \lambda {\}}$ of $gl(n)$ 
and its contragredient, provided that the anticommutator of odd 
generators closes on a \textit{polynomial} (of degree $>1$) in the 
even generators of $gl(n)$. A general discussion of admissible 
structure constants (\S 2) was exemplified by concrete fermionic 
oscillator constructions in specific cases (\S 3), and for
$gl_{2}(n/ {\{} 1 {\}} + {\{} \overline{1} {\}} )^{\texttt{a}, 
\alpha, \beta}$ (quadratic 
superalgebras with odd generators in the defining $n$-dimensional 
representation of $gl(n)$) a unique set of structure constants 
presented (\S 4), together with the elements of the construction of 
finite-dimensional irreducible representations.

As discussed in \S 1, our polynomial superalgebras are allied to 
classes of nonlinear algebras already studied in various physical 
settings.  For example, polynomial deformations of $sl(2)$ of degree 
$\Delta$ including the so-called Higgs\cite{Higgs} case ($\Delta = 2$) 
have been studied by Beckers\cite{Beckers}; nonlinear extensions of 
supersymmetric quantum mechanics have been identified in 
\cite{Debergh}.  Recently a broad class of `polynomial Lie algebras' 
has been found\cite{Andreev} in the context of anharmonic interactions 
in second quantised descriptions of many body systems.  In 
\cite{Andreev} the relationship of such models to integrable systems 
is studied.  In \cite{Karassiov}, various examples of polynomial Lie 
algebras were identified via their bosonic oscillator realisations.  
There the abstract status of such nonlinear algebras was not taken 
up to the extent of systematic detailed study of admissible Jacobi 
identities and structure constants, as in the present work for the 
super case.  However, many interesting variants of nonlinear 
polynomial algebras were obtained in oscillator constructions, 
including a classification of the three dimensional case (polynomial 
deformations of $su(2)$ and $su(1,1)$), and results on 
representations.  Analogous considerations for the $su(1,1)$ case are 
implicit also in recent work on `$K$-quantum' ladder operators and 
associated coherent and squeezed states \cite{An}.  In terms of 
classical and quantum dynamical systems, in \cite{Kalnins} various 
quadratic Poisson-Lie symmetry algebras have been investigated 
(together with quantum versions) in conection with various models of 
potentials admitting separation of variables.  Similarly the 
Askey-Wilson three-dimensional quadratic Poisson-Lie symmetry algebra 
(see for example \cite{Zhedanov} and references therein) plays a key 
role in non-linear integrable systems.  Finally, as noted in \S 1, the 
super-$gl(n)$ algebras are structurally closely related to the 
nonlinear $W$-algebras and $W$-superalgebras derived from the 
hamiltonian reduction on coadjoint Lie-Poisson manifolds which have 
been classified and studied in recent work 
\cite{Bouwknegt,Sorba93,Sorba96}.  Note that several examples of the 
latter which have been presented \cite{deBoer} can readily be 
transcribed into the covariant tensor notation used in the present 
paper.  For example, the $W$-algebra defined by the regular $sl(2)$ 
embedding within $sl(4)$, $\mbox{$\mathbf{4}$} \rightarrow 
\mbox{$\mathbf{2}$}_{0}+ \mbox{$\mathbf{1}$}_{1}+ 
\mbox{$\mathbf{1}$}_{-1}$ possesses an undeformed subalgebra $sl(2) + 
gl(1)$ with generators $E,F,H$ and $U$, together with a pair of 
doublets $\overline{G}_{\pm\frac12}, {G}_{\pm\frac12}$ with equal and 
opposite charge under $U$.  Defining (see (\ref{eq:CurlyEDefn}))
\[
\left( \begin{array}{cc} {E^{1}}_{1} & {E^{1}}_{2} \\
                          {E^{2}}_{1}& {E^{2}}_{2} \end{array} \right)
                          =
-\left( \begin{array}{cc} U+H & F \\
                          E & U-H \end{array} \right) ,
                          \]
the quadratic closure relations (8.4) of \cite{deBoer} read
\[
{[} \overline{G}^{a}, G_{b} {]} =  a{(E^{2})^{a}}_{b} +  b{E^{a}}_{b}+ c {\delta^{a}}_{b} \one 
\]                                            
in complete analogy with (\ref{eq:AntiCommutatorQbarQUnique}).
The study of the relationship between the present polynomial 
superalgebras, and hamiltonian reduction, is thus likely to shed light 
on their geometrical interpretation, in relation to their applications 
in quantum field theory (see below).  In particular, the meaning of 
`finite' symmetry transformations associated with deformed algebras in
general remains elusive.
 
In recent work on non-perturbative aspects of 
gauge field theories, the structure of the observable algebra has been investigated within 
the hamiltonian formulation on a finite lattice (see 
\cite{spinQED,scalQED,QCD1,QCD2,QCD3}), and has led to the 
need to study polynomial Lie superalgebras as a natural part of the 
algebraic structure. For completeness we briefly mention the context 
of these investigations, in order to explain the connection with the present work.
  
By definition, the observable algebra is the algebra of gauge invariant elements built from 
field operators, satisfying the Gauss law. 
In a first step, this algebra can be explicitly characterized in terms of generators
and relations. Next, it has to be endowed with an appropriate functional analytic structure and,
finally, one has to classify its irreducible representations. For quantum electrodynamics this 
programme has been implemented completely. It turns out that the observable algebra naturally 
decomposes into a bosonic part, which is isomorphic to a Heisenberg algebra of canonical 
commutation relations, and a matter field part.
For the case of spinor electrodynamics\cite{spinQED}, the matter field part turns out to 
be generated by the Lie algebra $u(2N)$, with $N$ denoting the 
number of lattice sites. For scalar electrodynamics\cite{scalQED}, it is generated by  
$u(N,N)$. In both cases, irreducible representations are labelled by the total electric charge, 
yielding a decomposition of the physical Hilbert space into charge superselection sectors.

In the case of quantum chromodynamics (QCD), a full analysis of the 
structure of the observable algebra is much more complicated (see 
\cite{QCD1,QCD2,QCD3}).  Here, quark matter fields are introduced as 
canonical fermionic operators 
${\psi^{*}}^a{}_{A}(\mbox{$\mathbf{x}$})$, 
$\psi_a{}^{A}(\mbox{$\mathbf{y}$})$, with $a = 1,2 \ldots, s$ spin 
$\mbox{}_{\alpha}, \mbox{}^{\dot{\alpha}} = 1,2$, and flavour indices, 
while $A, B = 1,2,3$ are $su(3)$ colour indices and $\mathbf{x}$, 
$\mathbf{y}$ $ = 1,2,\ldots, N$ are lattice sites with $N=L^{D}$ for a 
cubic lattice in dimension $D$.  Writing $a \mbox{$\mathbf{x}$} \equiv 
i, b \mbox{$\mathbf{y}$} \equiv j,\ldots$, natural colour invariant 
operators built from quark fields are then
\begin{eqnarray}
\label{observableJ}
E^i_{\gamma, j} & := & {\psi^{*i}}_{A} \, U^A_{\gamma \, B} \,
\psi_j{}^{B} \, \\
\label{observableW}
W_{\alpha \beta \gamma, (i j k)} &:= &
\frac{1}{6} \, \epsilon_{ABC} \, U^A_{\alpha \, D} \, U^B_{\beta \, E} \,
U^C_{\gamma \, F} \, \psi_i{}^{D} \,\psi_j{}^{E} \,\psi_k{}^{F} \, ,
\end{eqnarray}
with $U^A_{\gamma \, B} $ denoting the parallel transporter along $\gamma$, built from
the gluonic gauge fields. In formula (\ref{observableJ}), $\gamma$ denotes an arbitrary 
curve from $\mathbf{x}$ to $\mathbf{y}$, whereas in (\ref{observableW}) $\alpha$,
$\beta$ and $\gamma$ are arbitrary curves starting at some reference point
$\mathbf{t}$ and ending at $\mathbf{x}$, $\mathbf{y}$ and $\mathbf{z}$, respectively. 
The invariant operators $E^i_{\gamma, j}$ and $W_{\alpha \beta \gamma, (i j k)}$
represent hadronic matter of mesonic and baryonic type. 
These elements, together with a set of purely gluonic invariants\cite{QCD2,QCD3} 
constitute a set of generators. 
This set, however, is highly redundant. There is a number of non-trivial relations between
generators, inherited from the canonical (anti)- commutation
relations and from the local Gauss laws. A complete discussion of
the observable algebra as an abstract algebra in terms of
generators and defining relations will be presented in \cite{QCD3}. 

A method for solving a large part of the relations consists in choosing a lattice tree. 
Suppose that a tree has been fixed. Restricting 
ourselves then to invariant operators (\ref{observableJ}) and (\ref{observableW}), with 
$\alpha$, $\beta$ and $\gamma$ being the unique on-tree paths and the reference point being the
lattice root, these operators coincide with generators 
$\left( {E^{i}}_{j}\, , \, W_{(ijk)} \, , \, \overline{W}\mbox{}^{(ijk)} \right)$
of the algebra $gl_{2}(n/{\{}{3} {\}} + {\{} \overline{{3}} {\}} )$ discussed in \S 3. 
A slightly delicate gauge orbit analysis, together with some further tree techniques, enables 
one to further reduce the number of generators, leading to the algebra
$gl_{2}(n/{\{} 1^{3} {\}} + {\{} \overline{1^{3}} {\}} )$, defined in \S 3, with $gl(n)$-generators
given by formula (\ref{eq:EijDefn}) and
odd generators given by (\ref{eq:QbarQDefns}).
These generators still inherit some relations, but now the algebra has become tractable.
It can be shown that it is isomorphic to the universal enveloping algebra of 
$sl(N/1)$, factorized by a certain ideal defined in terms of relations on Casimir operators 
of a certain ordinary Lie superalgebra. Moreover, it is then easy to prove that 
irreducible representations of this algebra are labelled by the global colour charge (triality), 
built from the local colour charge densities carried by the quark field\cite{QCD2,QCD3}.

One aspect of the structure of `composite operators' such as 
the colour singlets $\overline{W}\mbox{}^{(ijk)}$ which emerges from 
the nonlinear algebra perspective has potentially wide applicability. Consider the reduction of degeneracy associated with 
states obtained by acting with monomials in $\overline{W}\mbox{}^{(ijk)}$ on the vacuum, 
relative to what would obtain if the $\overline{W}\mbox{}^{(ijk)}$, $W_{(pqr)}$ were elementary 
fermions (see \S A.3).  In the context of the representation 
theory of the polynomial super-$gl(n)$ algebras, the reduced representation 
content is a natural consequence of the Fock space realisation being 
generically an \textit{atypical} representation.

`Gauge invariance' is often handled by covariant BRST methods, which 
circumvent noncovariant hamiltonian approaches.  However, as a purely algebraic 
problem, the Gauss' law constraint can be also introduced via cohomology in hamiltonian 
BRST formulations.  To this end the equicovariant formalism of 
\cite{vanHolten} should be noted, wherein baryonic colour singlets such as 
(\ref{observableW}) are naturally identified as nontrivial cocycles 
at nonzero (but triality zero) ghost number. 

Applications of polynomial superalgebras in quantum field theory 
relate to spacetime supersymmetry.  In hamiltonian lattice QCD, the 
quadratic superalgebra is a \textit{bona fide} fermion-boson 
supersymmetry between baryon and meson states\footnote{Free field 
current algebras of this type within relativistic spin-flavour 
symmetry models were considered by Delbourgo, 
Salam and Strathdee\cite{DelbourgoSalam}.}.  There is the possibility that `no-go' 
theorems for the combination of internal and spacetime symmetries -- 
circumvented for supersymmetry to the extent of allowing $N$-extended 
Poincar\'{e} fermi-bose supersymmetries\cite{Haag} -- can be further 
relaxed for nonlinear supersymmetries.  Also, in the case $n=4$, 
appropriate real forms of $gl_{2}(4/ \{ 1 \} + \{ \overline{1} \} )$ 
may allow various six-dimensional realisations, or even new types of 
conformal supersymmetry in four dimensions ($su(2,2) \simeq so(4,2)$).  
Such superalgebras therefore add further to the resource of available 
generalised supersymmetries in diverse dimensions, following for 
example \cite{Nahm78}, or recently \cite{Sezgin98} for new higher 
dimensional superstring and supermembrane 
algebras\footnote{Parafermionic generalisations of Poincar\'{e} 
supersymmetry have also been considered, see \cite{Jarvis78}.}.

In relation to conformal and other space-time symmetries, it should be 
noted that antecedents of our polynomial algebras and superalgebras 
have been encountered before in connection with representation theory.  
For example Barut and Bohm\cite{BarutBohm} identify certain so-called 
special `representation relations' which are \textit{anticommutation} 
relations between the standard generators $P_{\mu}$ and $K_{\mu}$ and 
the Lorentz and dilatation generators $J_{\mu \nu}$ and $D$ of the form,
\[
{\{} P_{\mu} , K_{\mu} {\}} = (a \one + b D)\eta_{\mu \nu} +
{\{} {J_{\mu}}^{\sigma}, {J_{\sigma}}_{\nu}{\}}
\]
which obtain in certain classes of representation of the 
four-dimensional conformal algebra $so(4,2)$.  Similarly 
Bracken\cite{Bracken} in studying algebraic properties of the 
Gel'fand-Yaglom matrices $\Gamma_{\mu}$ in higher spin wave equations 
introduced analogous algebraic relations (but without the $D$ term) for ${\{} \Gamma_{\mu} , \Gamma_{\nu} {\}}$
as a generalisation of the Dirac algebra\footnote{The Dirac algebra 
itself is of course a `polynomial superalgebra' of degree zero!}. From the perspective of the 
present work, the above relations provide 
instances of the structure constants of quadratic superalgebras, in 
this case of the $so_{2}(3,1/{[}1{]})$ type, 
where the `odd' generators transform as vector operators. In contrast to the 
original contexts, however, any \textit{Lie algebra} 
relations which such vector operators happen to satisfy are now relegated 
to the status of specific `representation relations', with the 
\textit{anticommutation} relations regarded as primary.

In concrete applications the general question arises of a 
representation theory for the new polynomial superalgebras in their own right.  
In particular the existence of a tensor category associated with 
coproduct and Hopf structures needs further investigation, and the role 
of `deformations' needs clarification.  In the lattice QCD case, there 
is also the possibility that interesting structures may emerge only as 
\textit{local} entities in the thermodynamic ($N \rightarrow \infty$) 
limit.  Similarly, in the spacetime supersymmetry case, the 
appropriate context may be contraction limits of radii of additional 
dimensions, or orbifold parameters such as brane tensions.

In \S\S 3, 4 above, examples were found of polynomial superalgebras in 
which the Jacobi identities are underwritten by additional covariant 
Serre-type relations in the enveloping algebra.  Such identities are 
likely to be the rule rather than the exception for the nonlinear 
case, and may profoundly affect representations.  A precedent for such 
phenomena exists in the so-called `multiplet shortening' for massless 
supermultiplets in $N$-extended supersymmetries\cite{Strathdee}.  For 
appropriate kinematical conditions, the (spinor) supercharges are 
subjected to covariant constraints of the form 
$P_{\mu}{{\gamma^{\mu}}_{\alpha}}^{\beta}Q_{\beta} =0$.  In the usual 
Wigner induced representation method, this situation is easily handled 
as the representations of the abelian translation part are 
one-dimensional, and moreover $P_{\mu}$ is taken in a standard Lorentz 
frame.  In the $gl_{2}(n/ \{ 1 \} + \{ \overline{1} \} )$ and other 
cases, the constraints are also of covariant form, 
${E^{a}}_{b}\overline{Q}\mbox{}^{b} \propto \overline{Q}\mbox{}^{a}$, 
$Q_{a}{E^{a}}_{b} \propto Q_{b}$, but of course the multiplier 
${E^{a}}_{b}$ is nonabelian (for $n=4$, if $sl(4)$ can be identified 
with the real form $so(4,2)$, with $\overline{Q}, Q$ identified with 
the spinor representation of the latter, the above $\gamma \cdot P$ 
term would certainly appear as one contribution to the constraint).  
In this connection methods developed in recent 
work\cite{Gould,GouldBrackenHughes,JarvisGouldBracken} for the 
explicit construction of all (including atypical) finite-dimensional 
irreducible modules of type I Lie superalgebras can be adapted, at 
least for a class of representations of polynomial Lie superalgebras.  
In the ordinary Lie superalgebra case\cite{Gould}, the polynomial 
identities satisfied by the $gl(n)$ Gel'fand generators play a crucial 
role in deriving $gl(n)$ branching rules at each `floor' of the Kac 
module ${[} \sum_{k=0}^{n} \oplus \bigwedge^{k}(L_{-}){]} 
\otimes_{U_{+}} V(\Lambda)$.  The same method can be generalised to 
the polynomial super-$gl(n)$ case; the implications of the additional 
covariant Serre-type relations as constraints on the structure of the 
induced modules can be ascertained in precisely this 
framework\footnote{In the tensor operator language\cite{BrackenGreenJMP}, relations such as
${E^{a}}_{b}\overline{Q}\mbox{}^{b} = 
q \overline{Q}\mbox{}^{a}$, $ Q_{a}{E^{a}}_{b} = Q_{b} q $ in the 
$gl_{2}(n/{\{}1{\}}+{\{}\overline{1}{\}})$ case (and 
generalisations to $gl_{k}(n/{\{}\lambda{\}}+{\{}\overline{\lambda}{\}})$, see \S A.4) imply 
that $\overline{Q}, Q$ are shift operators for certain states in the 
$gl(n)$ modules ${\{}\lambda{\}}$,${\{}\overline{\lambda}{\}}$ 
depending on the eigenvalue $q$.}.  Full 
analysis along these lines, especially for the polynomial 
superalgebras related to hamiltonian lattice QCD, is the subject of 
future development.

\subsubsection*{Acknowledgements} 
The authors are grateful to J ~Kijowski for joint work, and for their
ongoing collaboration which provided the inspiration for the present
paper.  They acknowledge the generosity of the Alexander von Humboldt
Foundation, for providing Fellowship support to PDJ during leave at
the Institute for Theoretical Physics, University of Leipzig, and for
a reciprocal \textit{sur place} travel grant to GR for a visit to the
theory group, School of Mathematics, Physics and IASOS, University of
Tasmania.  The authors thank the respective host institutes for their
kind hospitality during these visits.  The research was partially
supported under the Australian Research Council Discovery project
DP-0208808, by the Institute for Theoretical Physics, University of
Leipzig, and by the Max-Planck Institut f\"{u}r Mathematik in den
Naturwissenschaften, Leipzig.  The authors thank the Centre for
Mathematical Physics, University of Queensland and its members for
hospitality during visits, in particular Mark Gould for discussions,
and pointing out the importance of Joseph's theorem, and Tony Bracken
for constructive comments on a draft version and for pointing out
references \cite{BarutBohm, Bracken}.  PDJ dedicates this work to the
memory of Professor H S (Bert) Green (1920-1999), mentor and personal
inspiration.

 %
 %
\begin{appendix}

\section*{Appendix}
\renewcommand{\theequation}{A.\arabic{equation}}
\setcounter{equation}{0}

\subsection*{A.1 Partition labelling for irreducible representations 
of $gl(n)$, and structure 
constants of polynomial super-$gl(n)$ algebras.}

The use of tensor notation has been formalised
by Weyl\cite{WeylBook}, Hamermesh\cite{HamermeshBook} and others in treatises on the relation between
partitions and irreducible finite-dimensional representations of the
group $GL(n)$.  A central role in the character theory is played by
the Schur functions, especially as developed by 
Littlewood\cite{LittlewoodBook} for
the unitary group $U(n)$ and subgroups $SU(n)$, $O(n)$ and $Sp(n)$. 
Many aspects of the theory have been developed further for arbitrary
semisimple (including exceptional) Lie groups, culminating in the
extensive tabulations of 
\cite{BlackKingWybourne83}\footnote{Supersymmetric partition labelling and Young
diagrams have also been introduced for representations of classical
superalgebras; see for example \cite{JarvisGreen79,DondiJarvis80, 
DondiJarvis81}. For an example of the organising power of group 
methods in tensor notation (applied to higher order heat 
kernel coefficients in curved space backgrounds) see for 
example\cite{Fulling}. See \cite{JarvisYung93a,JarvisYung93b} for applications 
of supersymmetric Schur functions to infinite-dimensional algebras.}.  
Most of the algorithms have been implemented in the
group theory package $^{\mbox{\copyright}}$\texttt{SCHUR}\cite{SCHUR}.  Here we outline
the necessary elements of the formalism for the case of
finite-dimensional irreducible representations of $GL(n)$ required for
the computation of structure constants (\S 1) and branching rules (\S
3 and \S A.2 below).

Finite dimensional irreducible representations of $GL(n)$ 
(corresponding to dominant integral highest weight modules of the simple complex 
Lie algebra $sl(n)$) are labelled\footnote{In Littlewood's 
nomenclature\cite{LittlewoodBook,BlackKingWybourne83} the symbols 
${\{}\lambda {\}}$, ${[}\lambda{]}$, ${<}\lambda{>}$ pertain to 
$GL(n)$, $O(n)$ and $Sp(n)$ repsectively.} by partitions ${\{} \lambda {\}} =
{\{} \lambda_{1}, \lambda_{2}, \ldots, \lambda_{\ell} {\}}$ for 
non-negative integers $\lambda_{1} \ge \lambda_{2} \ge \ldots \ge 
\lambda_{\ell} \ge 0$, where $\ell \le n$ is the number of parts of 
${\{}\lambda{\}}$, and ${\{}\lambda{\}}$ has weight or rank
$|\lambda| = \lambda_{1}+\lambda_{2}+\ldots+\lambda_{\ell}$; 
$\lambda$ is a partition of $|\lambda|$, $\lambda \vdash |\lambda|$.
${\{}\lambda {\}}$ is represented graphically by a Young  
tableau, which is an array of left-top justified rows of boxes, of 
lengths corresponding to the parts of ${\{}\lambda{\}}$.

Tensor or Kronecker products of the modules ${\{}\lambda {\}}$ and 
${\{}\mu {\}}$ are evaluated by the celebrated Littlewood-Richardson 
rule, which gives the resolution of the product of the corresponding 
characters (referred to as \textit{outer} multiplication of Schur 
$S$-functions),
\begin{equation}
{\{}\lambda {\}} \cdot {\{}\mu {\}} = \sum_{\nu} {C_{\lambda 
\mu}}^{\nu} {\{}\nu {\}}
\label{eq:LittlewoodRichardson}
\end{equation}
via the Littlewood-Richardson coefficients ${C_{\lambda \mu}}^{\nu}$, 
where $|\nu| = |\lambda|+|\mu|$.  Dually related is the definition of 
$S$-function \textit{skew},
\begin{equation}
{\{}\nu /\lambda {\}} = \sum_{\mu} {C_{\lambda \mu}}^{\nu} {\{}\mu 
{\}},
\label{eq:Skew}
\end{equation}
where the sum is over all ${\{}\mu {\}}$ such that ${\{}\lambda {\}} 
\cdot {\{}\mu {\}}$ $\ni {\{}\nu{\}}$, with the coefficient in the 
skew being given by the appropriate product multiplicity.

Various other $S$-function operations are needed for the manipulation 
of group representations and branching rules. Important is the \textit{inner} 
$S$-function product, defined for partitions of equal rank, $ |\lambda| = |\mu|$, 
\[
{\{}\lambda {\}} \circ {\{}\mu {\}} = \sum_{\nu} 
{\Gamma_{\lambda \mu}}^{\nu} {\{}\nu {\}},
\]
where $ |\nu| = |\lambda| = |\mu|$, which gives the resolution of the
corresponding product of characters (tensor product of
representations) of the symmetric group.  Finally there is the
$S$-function \textit{plethysm} ${\{}\lambda {\}} \otimes {\{}\mu{\}}$
which resolves the tensor product of the irreducible representation
${\{}\lambda {\}}$ with itself, $|\mu|$ times, into its projection of
symmetry type $\mu$, with respect to the action of the permutation
group $S_{|\mu|}$ on the factor spaces.

The manifest advantage of partition notation, even with its more
complicated extension to orthogonal and symplectic groups and even
exceptional groups (see \cite{BlackKingWybourne83}), is its general
feature of being \textit{rank-independent} for groups of large enough
dimension, and generic representations.  Any corrections for specific
groups (such as say $SU(3)$ or $SO(10)$), are done by means of
group-dependent \textit{modification rules} which rule out illegal
partitions resulting from general algorithms, and in some cases relate
characters specified by non-standard partitions to standard ones, up
to signs (which must be collected at the end of a calculation).

The major modification rule for $GL(n)$ is simply that partitions of 
more than $n$ parts (diagrams with more than $n$ rows) vanish 
identically.  In addition, for $SL(n)$, columns of length $n$ can be 
deleted.  When dealing with both covariant and contravariant 
representations of $GL(n)$, it is natural to introduce a more flexible 
mixed or composite partition notation ${\{}\overline{\lambda};\mu 
{\}}$ which represents a tensor of mixed contravariant and 
covariant rank $|\lambda|$, $|\mu|$ respectively, but for which all 
tensor contractions between upper and lower indices vanish\cite{King75}.  Standard 
partitions of this type have \textit{total} number of parts (rows of 
$\lambda$ and $\mu$) at most $n$, and are equivalent to canonical pure 
covariant or pure contravariant irreducible representations up to 
powers of the one dimensional alternating character (the determinant).  
Non-standard partitions of mixed type are either zero (for example if 
the number of parts is identically $n+1$) or modify in specific ways 
to standard tableaux.  We do not require the general rules (see 
\cite{BlackKingWybourne83}, and \cite{King75}), which have been 
implemented in $^{\mbox{\copyright}}$\texttt{SCHUR}\cite{SCHUR}.

The most obvious application of the composite 
notation is in handling the $n^{2}$-dimen\-sional adjoint representation 
of $GL(n)$. Technically this is isomorphic to the tensor product of 
the defining representation ${\{}1{\}}$ and its contragredient ${\{} 
\overline{1} {\}}$, written as
\[
{\{} \overline{1} {\}}\cdot {\{}1{\}} = {\{} \overline{1} ; 1{\}} + 
{\{}0{\}},
\]
with ${\{} \overline{1} ; 1{\}}$ representing the traceless part (the
adjoint representation of $SL(n)$), and ${\{}0{\}}$ the trace (the
linear Casimir invariant).  See \S 4, (\ref{eq:DetailedCR}) for the
explicit reduction.  For the familiar case of $SU(3)$, the composite
notation is convenient, in that all finite dimensional irreducible
representations can be specified by symmetrised tensors of mixed type. 
For example, ${\{} \overline{2} {\}} \equiv {\{}2^{2} {\}} \equiv
\mathbf{\overline{6}}$, ${\{} \overline{2};2 {\}} \equiv {\{}4,2 {\}}
\equiv \mathbf {27}$, and the product
\[
{\{} \overline{2} {\}}\cdot {\{} {2} {\}} = {\{} \overline{2};2 {\}}+ 
{\{} \overline{1};1 {\}}+  {\{} {0} {\}} 
\]
corresponds to the reduction
\[
\mathbf{\overline{6} \times 6 = 27 + 8 + 1.}
\]

We now formalise the above and other required tensor product and 
branching rules required in \S \S 2, 3.
For the tensor product of contravariant and covariant irreducible 
representations we have
\begin{equation}
	{\{} \overline{\lambda} {\}}\cdot {\{} {\mu} {\}} = \sum_{\alpha} {\{} 
	\overline{\lambda/\alpha} ; {\mu/\alpha} {\}}
	\label{eq:MixedProduct}
\end{equation}
where the skew is performed with respect to all legal 
${\{}\alpha{\}}$, and the expansion of the skew via (\ref{eq:Skew}) 
is done distributively.

For the decomposition of an irreducible representation of $GL(pq)$ 
with respect to $GL(p)\times GL(q)$ we have
\begin{equation}
	{\{}\lambda{\}} \downarrow \sum_{\sigma \vdash |\lambda|} {\{} 
	\sigma {\}} \times {\{} \sigma \circ \lambda {\}}.
\label{eq:GlpqBranching}
\end{equation}
In particular, if ${\{}\lambda{\}}= {\{}\ell{\}}$, then ${\{} \sigma \circ 
\lambda {\}}$ $= {\{} \sigma  {\}}$ as ${\{}\ell{\}}$ labels the 
trivial representation of $S_{\ell}$. Alternatively if ${\{}\lambda{\}}= 
{\{}1^{\ell}{\}}$, then ${\{} \sigma \circ 
\lambda {\}}$ $= {\{} \sigma' {\}}$, the partition transpose to ${\{} \sigma' 
{\}}$ (with rows and columns interchanged) as ${\{}1^{\ell}{\}}$ is 
the one-dimensional alternating character of $S_{\ell}$.

The rules for symmetric function plethysm have been developed 
by Littlewood (see appendix to \cite{LittlewoodBook}) and others; see 
for example \cite{ButlerWybourne,Chen84,Carvalho2001}. 
The algorithm for plethysm is implemented in the group theory 
package $^{\mbox{\copyright}}$\texttt{SCHUR}\cite{SCHUR}. However, for the applications 
needed in \S \S 2 and 4 above, the following rules suffice 
for the evaluation of low-rank cases\footnote{Note that the 
alternating signs in the final expression are typical of the outcome of 
Schur function manipulations, where a final positive sum of characters 
is only apparent after modification rules and cancellations have been 
accounted for.}:
\begin{eqnarray}
{\{} \overline{1} {\}} \cdot {\{} 1{\}}\otimes {\{} \ell {\}} &=& 
\sum_{\sigma \vdash \ell} {\{} \overline{\sigma}{\}} \cdot 
{\{}\sigma{\}} \nonumber \\
 &=& \sum_{\sigma \vdash \ell} \sum_{\alpha} {\{} 
 \overline{\sigma/\alpha } ; \sigma / \alpha {\}},
 \label{eq:AdjointPlethysmRule}
\end{eqnarray}
whereas for the irreducible part
\begin{eqnarray}
{\{} \overline{1} ; 1{\}}\otimes {\{} \ell {\}} &\uparrow& ({\{} 
\overline{1} {\}}\cdot {\{}1{\}} - {\{} {0} {\}} ) \otimes 
{\{}\ell{\}} \nonumber \\
&=& \sum_{m}(-1)^{m}({\{} \overline{1} {\}}\cdot 
{\{}1{\}})\otimes{\{}\ell-m{\}} \cdot {\{}0{\}} \otimes {\{}m{\}} 
\nonumber \\
&=& \sum_{m}(-1)^{m} \sum_{\sigma \vdash \ell-m} {\{}\overline{\sigma}{\}}\cdot{\{}\sigma{\}} 
\; \downarrow \; \sum_{m}(-1)^{m} \sum_{\sigma \vdash \ell-m} \sum_{\alpha} 
{\{}\overline{\sigma/\alpha} ; \sigma/\alpha{\}}.
\end{eqnarray}
From (\ref{eq:AdjointPlethysmRule}) it is evident in comparison 
with (\ref{eq:TensorCouplingsResults}) and (\ref{eq:LittlewoodRichardson}) 
that
\begin{eqnarray}
{n^{k}}_{\mu \nu} &=& \sum_{\alpha \vdash k, \gamma} {C_{\gamma \mu}}^{\alpha} {C_{\gamma 
\nu}}^{\alpha}, \nonumber \\
\quad {n^{\lambda}}_{\mu \nu} &=& \sum_{\gamma} {C_{\gamma 
\mu}}^{\lambda} {C_{\gamma 
\nu}}^{\lambda},
\end{eqnarray}
from which ${n^{k}}_{\mu \nu} \ge {n^{\lambda}}_{\mu \nu}$ provided $k 
\ge \ell$.

\subsection*{A.2 Generalised Gel'fand notation for $gl(n)$ defining
relations, and structure constants of polynomial super-$gl(n)$
algebras.}

We reiterate briefly here for the case of $gl(n)$, a framework for the
theory of characteristic identities for semisimple Lie algebras, which
puts the Gel'fand notation for the defining relations in a broader
context, and has been used in an essential way for the resolution of
the structure of atypical modules of type I classical superalgebras. 

Taken as a whole, within a certain (irreducible) representation, the
array of Gel'fand generators ${E^{a}}_{b}, 1 \le a,b \le n$ can be
regarded as an invariant ${E} \in \pi(gl(n)) \otimes
\mbox{End}({\mathbb C}^{n})$, where ${\mathbb C}^{n} \equiv V{
{\{}1{\}} }$ is the irreducible $n$-dimensional defining representation,
and $\pi : gl(n) \rightarrow \mbox{End}({\mathcal V})$ is an algebra
homomorphism for some $gl(n)$-module ${\mathcal V}$.  The
corresponding degree $k$ invariants ${E}^{k}$ within
$\pi(U(gl(n))) \otimes \mbox{End}(V{{\{}1{\}}})$, are nothing but the
above matrix powers ${(E^{k})^{a}}_{b}$ of the array of Gel'fand
generators; the traces $ \langle E^{k} \rangle$ are of course the
standard Casimir operators of $gl(n)$. 

This construction generalises
to an invariant ${\mathcal E} \in \pi(gl(n)) \otimes
\mbox{End}(V{{\{}\lambda{\}}})$ for an arbitrary irreducible
representation ${\{}\lambda {\}}$.  The matrix elements with respect 
to a basis of $V{{\{}\lambda{\}}}$ of ${\mathcal E}^{k} \in \pi(U(gl(n))) \otimes
\mbox{End}(V{{\{}\lambda{\}}})$ will provide precisely the leading
unique degree $k$ coupling for the polynomial superalgebra, and
moreover related partial traces enable the remaining lower degree
couplings to be enumerated in accord with the above counting schemes.

Let $\texttt{C}$ be the second order Casimir invariant (see 
(\ref{eq:CasimirEvalues}) above).  The general definition of 
${\mathcal E}$ is:
\begin{equation}
{\mathcal E} = \textstyle{\frac 12} (\pi \otimes \one) \circ ( 
\Delta(\texttt{C}) - \texttt{C} \otimes \one - \one \otimes 
\texttt{C} ).
\label{eq:CurlyEDefn}
\end{equation}
Finally, if $e_{a}\otimes e_{b} \otimes \ldots$ are an (appropriately 
symmetrised) basis for $V{{\{}\lambda{\}}}$, then the set of $gl(n)$ 
generators in generalised Gel'fand notation is defined by the the 
matrix elements
\begin{equation}
{{\mathcal E}^{abc\ldots}}_{pqr\ldots} = (e_{a}\otimes e_{b} \otimes 
\ldots, {\mathcal E} e_{p}\otimes e_{q} \otimes \ldots )
\label{eq:GeneralisedGelfandBasis}
\end{equation}
as operators in $\mbox{End}({\mathcal V})$\footnote{The above 
formalism has been used to derive polynomial characteristic identities 
for generators of Lie algebras and superalgebras.  See 
\cite{BrackenGreenJMP} for original works (see also \cite{Okubo75}), and 
\cite{Edwards,JarvisGreen79} for extensions to superalgebras.  For 
abstract approaches see \cite{CantCareyHurst,Gould78,Gould87}.  For the 
relationship of the construction to Casimir invariants of arbitrary 
degree see \cite{Okubo,King}.  For the role of Yangians in relation to 
Laplace operators for Lie algebras and noncommutative characteristic 
polynomials see \cite{Molev}.  For the relation to the Goddard Kent 
Olive construction\cite{GoddardKentOlive} in the affine case see 
\cite{JarvisMcAnally}.}.

\subsection*{A.3 Decomposable representations of $gl_{2}(n/{\{}3{\}} + 
{\{}\overline{3}{\}})$.}

The examples of oscillator realisations which we have considered not 
only provide the defining relations of various types of polynomial 
super-$gl(n)$ algebras, but also furnish examples of representations.  
In the fermionic case there are thus finite-dimensional, generally 
decomposable, super-$gl(n)$ representations in Fock space, via the 
usual action, and in the associated Clifford algebra, via the adjoint 
action. In this appendix we study the case
$gl_{2}(n/{\{}3{\}} + {\{}\overline{3}{\}})$. The relevant state space 
and adjoint operators are classified in the general case, 
and for concreteness the results for the simplest cases $n=1$, $n=2$ are 
listed explicitly.

As discussed in \S 3 above, within the $m=6n$-dimensional Clifford
algebra generated by the fermionic creation and annihilation operators
$a^{iA}$, $a_{jB}$, $gl_{2}(n/{\{}3{\}} + {\{}\overline{3}{\}})$ is
generated by $\overline{W}\mbox{}^{(ijk)}$, $W_{(pqr)}$, and ${E^{i}}_{j}$ realised as
colour traces.  The total Fock space is as usual an irreducible spinor
representation of $so(6n) \supset gl(3n)$, and the Clifford algebra is
embedded naturally as all endomorphisms on this space; taking account
of the grading, therefore, we have $gl(n/{\{}3{\}} +
{\{}\overline{3}{\}})_{2} \subset C\ell(6n) \subset
gl(2^{3n-1}/2^{3n-1})$.

Given normal ordering conventions, and the usual construction of
states using creation modes applied to the vacuum state, the problem
of classifying colour singlet states in Fock space is in fact a
sub-case of that of identifying all colour-singlet operators.  In
general, we consider the reduction of the tensor representation of
$gl(3n)$, corresponding to the product
\[
{X^{i_{1}A_{1}i_{2}A_{2} \ldots i_{K}A_{K}}}_
              {p_{1}B_{1} p_{2}B_{2}\ldots p_{L}B_{L} } =
       a^{i_{1}A_{1}}a^{i_{2}A_{2}}\ldots 
       a^{i_{K}A_{K}}a_{p_{1}B_{1}}a_{p_{2}B_{2}}\ldots a_{p_{L}B_{L}}
\]
to $gl(3)+gl(n)$, with the identification of $sl(3)$ singlets (colour 
invariant \textit{states} correspond to the $L=0$ case)\footnote
{In the context of physical applications the relevant symmetry groups 
are in the unitary chain, and the colour transformations belong to 
$SU(3)$. Here we merely count singlets of $sl(3)$.}. This is 
a standard group reduction problem\footnote{Corresponding to the 
reduction of the spinor representation of $so(6n)$ under
branching chain $so(6n) \supset gl(3n) \supset gl(3)\!+\!gl(n)$}, and can be efficiently handled via 
the extended partition labelling (see \S A.1 above, and \cite{BlackKingWybourne83}), resulting in:
\begin{equation}
	gl(3n) \downarrow gl(3)+gl(n) \quad 
{\{}\overline{ 1^{K}} {\}}\cdot {\{} 1^{L}{\}} 
\downarrow \sum_{\rho \vdash K, \sigma \vdash L} {\{} \overline{\rho'} {\}}\cdot {\{} \sigma' {\}} \times
{\{} \overline{\rho} {\}}\cdot {\{} \sigma {\}},
\label{eq:Gl3GlnDecomp}
\end{equation}
where $0 \le K,L \le 3n$, and ${\{} \rho' {\}}$, ${\{} \sigma' {\}}$ 
are the transpose partitions to ${\{} \rho' {\}}$, ${\{} \sigma' {\}}$ 
such that in the permutation groups $S_{K}$, $S_{L}$ we have ${\{} 
\rho' \circ \rho{\}} \ni {\{} 1^{K} {\}}$, and similarly ${\{} \sigma' 
\circ \sigma{\}} \ni {\{} 1^{L} {\}}$ (see appendix A.1, and 
\cite{BlackKingWybourne83}).  Using the usual restrictions that 
partitions represent nonzero characters of $gl(n)$ provided they have 
at most $n$ rows, it can be seen from (\ref{eq:Gl3GlnDecomp}) that 
both $\rho$ and $\sigma$ must fall within a rectangular envelope of 
standard shape $3 \times n$.  Finally, the right hand side of 
(\ref{eq:Gl3GlnDecomp}) should be reduced with respect to $sl(3)$, 
which entails the further modification rule that columns of $\rho'$, 
$\sigma'$ of length three can be removed.  Thus for $K=0$, the 
branching rule gives an $sl(3)$ singlet provided ${\{} \sigma {\}} = 
{\{} 3^{r} {\}}$, $r=0,1,2\ldots,n$ (`multi-baryons'), and similarly 
for $L=0$ we have ${\{} \overline{\rho} {\}} = {\{} 3^{r} {\}}$, 
$r=0,1,2\ldots,n$ (`multi anti-baryons', respectively).  For both $K,L 
\ne 0$, we count $sl(3)$ colour singlets within ${\{} \overline{\rho'} 
{\}}\cdot {\{} \sigma' {\}}$ for legal partitions $\rho', \sigma'$ 
within the $n \times 3$ rectangle.  The relevant branching rule is 
(see appendix \S A.1 and \cite{BlackKingWybourne83}):
\[
{\{} \overline{\rho'} {\}}\cdot {\{} \sigma' {\}} \downarrow
\sum_{\alpha} {\{} \overline{\rho' / \alpha} ; \sigma' / \alpha {\}}
\]
where the sum is over all partitions $\alpha$ whose skew with both
$\rho'$, $\sigma'$ is nonvanishing.  Obviously, the sum contains a
singlet if and only if $\rho = \sigma$ (skewing by $\alpha = \rho =
\sigma$ leads to the trivial representation).  For $K=L$ this
immediately gives a classification of all `meson' colour singlets
within $gl(n)$, classified by the reduction of ${\{} \overline{\sigma}
{\}}\cdot {\{} \sigma {\}}$ for $\sigma$ within the standard $3 \times
n$ rectangular envelope.  For $K>L$ or $K<L$ the same theorem applies,
but the equality of $\rho$ and $\sigma$ may arise as a result of
modification by dropping columns of length three.  This provides a
classification of all `exotic baryons and anti-baryons', according to
$gl(n)$ multiplets arising in the reduction of ${\{}
\overline{3^{r}\wr \lambda} {\}}\cdot {\{} 3^{s} \wr \lambda {\}}$,
where $r \ne s$, $0 \le r,s \le n$, where $\lambda$ lies within a
truncated $2 \times (n-t)$ rectangular envelope, where
$t=\mbox{max}(s,t)$ and the notation ${\{} {3^{r}\wr \lambda} {\}}$
indicates that $\lambda$ is appended \textit{below} the relevant
rectangular block of depth $r$.  This classification in fact includes
the previous $K=0$ and $L=0$ cases, which appear as either $r = 0$,
$\lambda = \phi$, or $s=0$, $\lambda = \phi$, respectively.

The above colour singlet states and operators are easy to enumerate explicitly for the lowest                 
cases $n=1$ and $n=2$. For the former, from the general result of \S 3, the polynomial 
superalgebra is isomorphic to $gl(1/1)$, so the classification of 
states and operators can at the same time be viewed as a list of $gl(1/1)$ 
representations.  The colour singlet states and operators are given 
in table \ref{tbl:OperatorsGl1}.

For $n=2$, table \ref{tbl:OperatorsGl2} provides a list of colour 
$sl(3)$ singlet $gl(2)$ 
representations ${\{} \overline{\sigma} {\}}\cdot {\{} \sigma {\}}$,
${\{} \overline{\rho} {\}}\cdot {\{} \sigma {\}}$, for meson, 
baryon, exotic baryon and dibaryon operators, and conjugates. From the partitions, it is easy 
to reconstruct the parameters $r$, $s$ and $\lambda$ used above; the fermion 
content $(K,L)$ of each state or operator is given, together with the 
$sl(2)$ spin content (written as a reducible representation 
$j\times k$  corresponding to the reduction of ${\{} \overline{\rho} 
{\}}$ and ${\{} \sigma {\}}$ respectively), together with the  
dimension $N= (2j+1)(2k+1)$.

\subsection*{A.4 Relation between  $gl_{2}(4/{\{}1^3{\}} + 
{\{}\overline{1^3}{\}})$ and $gl_{2}(4/{\{}1{\}} + 
{\{}\overline{1}{\}})^{\texttt{a},\alpha,\beta}$.}

The approach of \S 4 was to analyze abstractly the defining relations
of the $gl_{2}(n/{\{}1{\}} + {\{}\overline{1}{\}})$ family 
of quadratic algebras in order to establish properties of the 
admissible structure constants in the absence of a particular 
realisation. Here we reverse this philosophy and show, for the
case $n=4$, the relation between the previously considered (fermionic 
oscillator) $gl_{2}(4/{\{}1^3{\}} + 
{\{}\overline{1^3}{\}})$ construction, and $gl_{2}(4/{\{}1{\}} + 
{\{}\overline{1}{\}})^{\texttt{a},\alpha,\beta}$.

Consider then the generators $\overline{Q}\mbox{}^{ijk}$, $Q_{pqr}$ 
of $gl_{2}(4/{\{}1^{3}{\}} + {\{}\overline{1^{3}}{\}})$ as in 
$(\ref{eq:SuperGl1^3})$, but with the modification that the odd 
generators transform as tensor densities of weight $w$,
\begin{eqnarray}
	{[}{E^{i}}_{j}, \overline{Q}\mbox{}^{[k \ell m]}{]} &=& 
	{\delta_{j}}^{k}\overline{Q}\mbox{}^{[i \ell m]} + 
	{\delta_{j}}^{\ell}\overline{Q}\mbox{}^{[k i m]} + 
	{\delta_{j}}^{m}\overline{Q}\mbox{}^{[k \ell i]} + w {\delta^{i}}_{j}\overline{Q}\mbox{}^{[k \ell m]} , \nonumber \\
	{[}{E^{i}}_{j}, {Q}_{[pqr]}{]} &=& 
	-{\delta^{i}}_{p} {Q}_{[jqr]} -{\delta^{i}}_{q}{Q}_{[pjr]} 
    -{\delta^{i}}_{r}{Q}_{[pqj]} - w {\delta_{i}}^{j} {Q}_{[pqr]}.	
		\label{eq:TensorDensities}
\end{eqnarray}
Then, defining 
\begin{eqnarray}
\overline{S}\mbox{}^{i}=\frac 16 \epsilon^{ijk\ell} Q_{jk\ell}, &\;& 
Q_{jk\ell} = - \epsilon_{jk\ell m} \overline{S}\mbox{}^{m}, \nonumber \\
S_{i} = \frac 16 \epsilon_{ijk\ell} 
\overline{Q}\mbox{}^{jk\ell} , &\;& \overline{Q}\mbox{}^{jk\ell} = - 
\epsilon^{ijk\ell}S_{\ell},
\label{eq:SSbarDefns}
\end{eqnarray}
produces 
\begin{eqnarray}
{[}{E^{i}}_{j}, \overline{S}\mbox{}^{k}{]} &=& 
{\delta_{k}}^{j}\overline{S}\mbox{}^{i} -(1+w) {\delta_{i}}^{j} 
\overline{S}\mbox{}^{k}, \\
{[}{E^{i}}_{j},S_{k} {]} &=& - {\delta_{i}}^{k} S_{j} + (1+w) {\delta_{i}}^{j}S_{k},
\label{eq:SSbarDensities}
\end{eqnarray}
so that the choice $w=-1$ leads to standard tensor transformation 
rules for the rank one odd generators $S$, $\overline{S}$. Proceeding 
with (\ref{eq:SSbarDefns}) produces
\[
{\{} \overline{S}\mbox{}^{i}, S_{j} {\}} =
\frac 16 {[} -3 {\{} \overline{Q}\mbox{}^{[ik \ell]}, Q_{[jk\ell]} 
{\}} + {\delta^{i}}_{j} {\{} \overline{Q}\mbox{}^{[k \ell m]}, Q_{[k \ell 
m]} {]},
\]
and use of the structure constants (\ref{eq:SuperGl1^3}) together with the 
standard definitions (\ref{eq:DetailedCR}) leads after use of 
(\ref{eq:AntiCommutatorQbarQ}) to the quoted form
\[
{\{} \overline{S}\mbox{}^{i}, S_{j}{\}} = {(J^{2})^{a}}_{b} -\frac 12 \langle 
	J^{2}\rangle {\delta^{a}}_{b} - \frac 12\widehat{N}{J^{a}}_{b}
	+ (\frac{3}{16}\widehat{N}^{2} -\frac 34 \widehat{N} + 2){\delta^{a}}_{b},
\label{eq:AntiCommutatorSbarS}
\]
which agrees with (\ref{eq:DetailedCR}), (\ref{eq:RawCoeffs}) for $n=4$
($a=-\frac 12$, $\texttt{b}_{1}=-\frac 12$, $\texttt{c}_{1}= 
\frac {3}{16}$), together with
$\texttt{b}_{2}=-1, \texttt{c}_{2}= -\frac 34$.

Finally we comment on the role of the covariant constraints 
(\ref{eq:ConstraintsJQbarQJ}) in this case, and generalisations to other 
cases. Firstly for the antisymmetric rank 3 case $gl_{2}(n/{\{}1^3{\}} + 
{\{}\overline{1^3}{\}})$, in the realisation 
(\ref{eq:SuperGl1^3}) via fermionic oscillator modes, we have directly 
from (\ref{eq:QbarQDefns}), (\ref{eq:EijDefn})  that
\begin{eqnarray}
{E^{i}}_{m} \overline{Q}\mbox{}^{[m j k]} + {E^{j}}_{m} 
\overline{Q}\mbox{}^{[m k i]} + {E^{k}}_{m} \overline{Q}\mbox{}^{[m i j]} 
&=& (-3 \widehat{N} +3(n+1)\one) \overline{Q}\mbox{}^{[i j k]}, 
\nonumber \\
Q_{[ijm]}{E^{m}}_{k} + Q_{[jkm]}{E^{m}}_{i} + Q_{[kim]}{E^{m}}_{j} &=&
Q_{[ijk]}(-3 \widehat{N} +3(n+1)\one).	
\label{eq:AntiSymmQbarQconstraints}
\end{eqnarray}
Similarly in the symmetric rank 3 case $gl_{2}(n/{\{}3{\}} + 
{\{}\overline{3}{\}})$ we have from (\ref{eq:ColourEijDefn}), (\ref{eq:WbarWDefns}) that
\begin{eqnarray}
{E^{i}}_{m} \overline{W}\mbox{}^{(m j k)} + {E^{j}}_{m} 
\overline{W}\mbox{}^{(m k i)} + {E^{k}}_{m} \overline{W}\mbox{}^{(m i 
j)} 
&=& (\widehat{N} -3(n-3)\one) \overline{W}\mbox{}^{(i j k)}, 
\nonumber \\
W_{(ijm)}{E^{m}}_{k} + W_{(jkm)}{E^{m}}_{i} + W_{(kim)}{E^{m}}_{j} &=&
W_{(ijk)}(\widehat{N} -3(n-3)\one).	
\label{eq:SymmWbarWconstraints}
\end{eqnarray}
which reflect the identity (true for any $su(3)$ tensor $T_{A}$)
\[
\epsilon_{ABC}T_{D}- 
\epsilon_{BCD}T_{A}+\epsilon_{CDA}T_{B}-\epsilon_{DAB}T_{C} =0.
\]
Note that the left-hand sides of (\ref{eq:AntiSymmQbarQconstraints}), 
(\ref{eq:SymmWbarWconstraints}) can be expressed in the form of the 
action of the invariants 
(\ref{eq:GeneralisedGelfandBasis}) on the appropriate odd generators 
(that is, the matrix action of the set of generalised Gel'fand basis 
generators of $gl(n)$ on $Q$, $\overline{Q}$, and $W$, 
$\overline{W}$), respectively. Finally for $S$, $\overline{S}$ as 
in (\ref{eq:SSbarDefns}), and ${E^{i}}_{j}$ defined as in 
(\ref{eq:EijDefn}), we have directly from (\ref{eq:SSbarDensities}) 
with $w=-1$, (\ref{eq:AntiSymmQbarQconstraints}) that
\begin{eqnarray}
	(E \cdot \overline{S})^{i} & = & (4 \widehat{N} - 
	15 \one) \overline{S}\mbox{}^{i}, \nonumber \\
	(S \cdot E)_{i} & = & S_{i}( 4 \widehat{N} - 15 \one)
    \label{eq:SbarSconstraints}
\end{eqnarray}
However, as is evident from (\ref{eq:RawCoeffs}), 
(\ref{eq:SerreCoeffs}), in the case $\texttt{a}=-\frac 12$, the 
structure coefficients are independent of the particular form of the 
constraint.

\begin{table}[tbp]
	\centering
	\caption{Colour singlet composite operators for 
	$gl_{2}(1/{\{}{3}{\}}+{\{}\overline{3}{\}}) \simeq gl(1/1)$, listed by fermion
	content $(K,L)$, and $gl(1)$ content
	${\{}\overline{K}{\}}\cdot{\{}L{\}}$ (one dimensonal 
	representations with $gl(1)$ quantum number $K-L$). 
	Mesons have $K=L$, baryons have $K=0$
	or $L=0$.  The
	 adjoint module has
	even dimension 4, and odd dimension 
	2 (including the 2 odd and 2 even generators of 
	$gl(1/1)$).}\vspace*{1cm}
		\begin{tabular}{|c||c|}
			\hline
			$(K,L)$ & ${\{}\overline{K}{\}}\cdot{\{}L{\}}$  \\
			\hline \hline
			\rule[-.6cm]{0cm}{1.2cm}$(0,0)$ & ${\{}\overline{0}{\}}\cdot{\{}0{\}}$    \\
			\hline
			\rule[-.6cm]{0cm}{1.2cm}$(1,1)$ & ${\{}\overline{1}{\}}\cdot{\{}1{\}}$    \\
			\hline
			\rule[-.6cm]{0cm}{1.2cm}$(2,2)$ & ${\{}\overline{2}{\}}\cdot{\{}2{\}}$   \\
			\hline
			\rule[-.6cm]{0cm}{1.2cm}$(3,3)$ & ${\{}\overline{3}{\}}\cdot{\{}3{\}}$   \\
			\hline \hline
			\rule[-.6cm]{0cm}{1.2cm}$(0,3)$ & ${\{}\overline{0}{\}}\cdot{\{}3{\}}$   \\
			\hline
			\rule[-.6cm]{0cm}{1.2cm}$(3,0)$ & ${\{}\overline{3}{\}}\cdot{\{}0{\}}$   \\
			\hline \hline
			
		\end{tabular}

	\label{tbl:OperatorsGl1}
\end{table}

\begin{table}[tbp]
	\centering
	\caption{Colour singlet composite operators for 
	$gl_{2}(2/ {\{}{3}{\}}+{\{}\overline{3}{\}})$, listed by fermion
	content $(K,L)$, $gl(2)$ content
	${\{}\overline{\rho}{\}}\cdot{\{}\sigma{\}}$ for partitions in the
	standard $3 \times (n=2)$ rectangular envelope, and $sl(2)$ spin
	content $j \times k$ with dimension $N=(2j+1)(2k+1)$.  Mesons
	have $K=L$, baryons and dibaryons and their conjugates have $K=0$
	or $L=0$, and exotic baryons have $K > L > 0$ or $L > K >0$.  The
	 adjoint module has
	even dimension 52, and odd dimension 
	40 (including the 8 odd and 4 even generators).}\vspace*{1cm}
	\parbox[tbp]{15cm}{
		\begin{tabular}{|c||c|c|c|}
			\hline
			$(K,K)$ & ${\{}\overline{\sigma}{\}}\cdot{\{}\sigma{\}}$ & $j\times j$ & 
			$N$   \\
			\hline \hline
			\rule[-.6cm]{0cm}{1.2cm}$(0,0)$ & ${\{}\overline{0}{\}}\cdot{\{}0{\}}$ & $0 \times 0$ & 1   \\
			\hline
			\rule[-.6cm]{0cm}{1.2cm}$(1,1)$ & ${\{}\overline{1}{\}}\cdot{\{}1{\}}$ & $\frac 12 \times \frac 
			12$ & 4   \\
			\hline
			\rule[-.6cm]{0cm}{1.2cm}$(2,2)$ & ${\{}\overline{2}{\}}\cdot{\{}2{\}}$ & $1 \times 1$ & 9   \\
			
			 & ${\{}\overline{1^{2}}{\}}\cdot{\{}1^{2}{\}}$ & $0 \times 0$ & 1   \\
			\hline
			\rule[-.6cm]{0cm}{1.2cm}$(3,3)$ & ${\{}\overline{3}{\}}\cdot{\{}3{\}}$ & $\frac 32 \times \frac 
			32$ & 16   \\
			
			 & ${\{}\overline{2,1}{\}}\cdot{\{}2,1{\}}$ & $\frac 12 \times \frac 
			12$ & 4   \\
			\hline
		\rule[-.6cm]{0cm}{1.2cm}	$(4,4)$ & ${\{}\overline{3,1}{\}}\cdot{\{}3,1{\}}$ & $1 \times 1$ & 9   \\
			
			 & ${\{}\overline{2,2}{\}}\cdot{\{}2,2{\}}$ & $0 \times 0$ & 1   \\
			\hline
			\rule[-.6cm]{0cm}{1.2cm}$(5,5)$ & ${\{}\overline{3,2}{\}}\cdot{\{}3,2{\}}$ & $\frac 12 \times \frac 
			12$ & 4   \\
			\hline
			\rule[-.6cm]{0cm}{1.2cm}$(6,6)$ & ${\{}\overline{3,3}{\}}\cdot{\{}3,3{\}}$ & $0 \times 0$ & 1   \\
			\hline \hline
			
		\end{tabular}
\hspace{1cm}
		\begin{tabular}{|c||c|c|c|}
			\hline
			$(K,L)$ & ${\{}\overline{\rho}{\}}\cdot{\{}\sigma{\}}$ & $j\times k$ & 
			$N$   \\
			\hline \hline
			\rule[-.6cm]{0cm}{1.2cm}$(0,3)$ & ${\{}\overline{0}{\}}\cdot{\{}3{\}}$ & $0 \times \frac 32$ & 4   \\
			\hline
			\rule[-.6cm]{0cm}{1.2cm}$(3,0)$ & ${\{}\overline{3}{\}}\cdot{\{}0{\}}$ & $\frac 32 \times 0$ & 4   \\
			\hline \hline
			\rule[-.6cm]{0cm}{1.2cm}$(1,4)$ & ${\{}\overline{1}{\}}\cdot{\{}3,1{\}}$ & $\frac 12 \times 1$ & 6   \\
			\hline
			\rule[-.6cm]{0cm}{1.2cm}$(2,5)$ & ${\{}\overline{2}{\}}\cdot{\{}3,2{\}}$ & $1 \times \frac 12$ & 6   \\
			\hline
			\rule[-.6cm]{0cm}{1.2cm}$(3,6)$ & ${\{}\overline{3}{\}}\cdot{\{}3^{2}{\}}$ & $\frac 32 \times 0$ & 4   \\
			\hline
			\rule[-.6cm]{0cm}{1.2cm}$(4,1)$ & ${\{}\overline{3,1}{\}}\cdot{\{}1{\}}$ & $1 \times \frac 12$ & 6   \\
			\hline
			\rule[-.6cm]{0cm}{1.2cm}$(5,2)$ & ${\{}\overline{3,2}{\}}\cdot{\{}2{\}}$ & $\frac 12 \times 1$ & 6   \\
			\hline
			\rule[-.6cm]{0cm}{1.2cm}$(6,3)$ & ${\{}\overline{3^{2}}{\}}\cdot{\{}3{\}}$ & $0 \times \frac 32$ & 4   \\
			\hline \hline
			\rule[-.6cm]{0cm}{1.2cm}$(0,6)$ & ${\{}\overline{0}{\}}\cdot{\{}3^{2}{\}}$ & $0 \times 0$ & 1   \\
			\hline
			\rule[-.6cm]{0cm}{1.2cm}$(6,0)$ & ${\{}\overline{3^{2}}{\}}\cdot{\{}0{\}}$ & $0 \times  0$ & 1   \\
			\hline \hline
			
		\end{tabular}
	}	\label{tbl:OperatorsGl2}
\end{table}

\end{appendix}
\end{document}